\documentclass[pra, aps,letterpaper, twocolumn, preprintnumbers,superscriptaddress]{revtex4}

\usepackage{verbatim}
\usepackage{amsmath}
\usepackage{latexsym}
\usepackage{revsymb}
\usepackage{yfonts}

\usepackage{natbib}
\usepackage{amsfonts}
\usepackage{amsmath}
\usepackage{amssymb}
\usepackage{amsthm}
\usepackage{graphicx}
\usepackage{bm}
\usepackage{bbm}

\newcommand{\benn}{\begin{eqnarray*} \begin{aligned}}
\newcommand{\eenn}{\end{aligned} \end{eqnarray*} }

\newcommand{\balg}{\begin{align}}
\newcommand{\ealg}{\end{align}}
\newcommand{\bc}{\begin{center}}
\newcommand{\ec}{\end{center}}

\newcommand{\id}{\mathbb{I}}

\newcommand{\tr}{\mathop{\mathrm{tr}}\nolimits}

\newtheorem{theorem}{Theorem}[section]

\newtheorem{lemma}[theorem]{Lemma}

\newcommand{\hil}{\mathcal{H}}

\usepackage{amsfonts}

\def\id{\mathbb{I}}

\def\01{\{0,1\}}

\newcommand{\eps}{\varepsilon}
\newcommand{\ket}[1]{|#1\rangle}
\newcommand{\bra}[1]{\langle#1|}

\newcommand{\proj}[1]{|#1\rangle\langle#1|}

\newcommand{\inp}[2]{\langle{#1}|{#2}\rangle} 

\newcommand{\ii}{\ensuremath{{\rm I}}}
\newcommand{\bh}{\ensuremath{\breve{{\rm H}}}}
\newcommand{\ph}{\ensuremath{\widehat{{\rm H}}}}

\newcommand{\pht}{\ensuremath{\widehat{{\rm H}}_{+}}}

\newcommand{\sh}{\ensuremath{{\rm H}}}
\newcommand{\nh}{\ensuremath{{\rm S}}}
\newcommand{\sv}{\ensuremath{{\rm S}}}
\newcommand{\ih}{\ensuremath{\widehat{{\rm I}}}}

\newcommand{\iht}{\ensuremath{\widehat{{\rm I}}_{+}}}

\newcommand{\bop}{\mathcal{B}}
\newcommand{\cS}{\mathcal{S}}
\newcommand{\cR}{\mathcal{R}}
\newcommand{\ef}{\textbf{e}}
\newcommand{\fef}{\textbf{f}}
\newcommand{\hef}{\textbf{h}}
\newcommand{\gef}{\textbf{g}}

\newcommand{\xout}{x_{\rm out}}
\newcommand{\xin}{x_{\rm in}}
\newcommand{\vxout}{\vec{x}_{\rm out}}
\newcommand{\vxin}{\vec{x}_{\rm in}}
\newcommand{\yout}{y_{\rm out}}
\newcommand{\yin}{y_{\rm in}}
\newcommand{\zout}{z_{\rm out}}
\newcommand{\zin}{z_{\rm in}}

\newcommand{\para}[1]{\addtocounter{para}{1}\bigskip\noindent\emph{(\alph{para})$\;$ #1}.$\;$}
\newcounter{para}[section]

\newcommand{\h}{\ensuremath{\widetilde{{\rm H}}}}

\newcommand{\states}{\mathcal{S}}

\newcommand{\mbf}[1]{\mathbf{#1}}

\newcommand{\dist}{\mathcal{D}}
\newcommand{\cdist}{\mathcal{C}}

\newcommand{\cE}{\mathcal{E}}

\newcommand{\src}{\ensuremath{\textsf{Src}}}

\begin{document}

\title{Entropy in general physical theories}

\author{Anthony J. Short}
\affiliation{DAMTP, Centre for Mathematical Sciences, Wilberforce
Road, Cambridge CB3 0WA, UK} \email{ajs256@cam.ac.uk}
\author{Stephanie Wehner}
\affiliation{Institute for Quantum Information, Caltech, Pasadena, CA 91125, USA}
\email{wehner@caltech.edu}

\begin{abstract}
Information plays an important role in our understanding of the
physical world. We hence propose an entropic measure of information
for \emph{any} physical theory that admits systems, states and
measurements. In the quantum and classical world, our measure
reduces to the von Neumann and Shannon entropy respectively. It can
even be used in a quantum or classical setting where we are only
allowed to perform a limited set of operations. In a world that
admits superstrong correlations in the form of non-local boxes, our
measure can be used to analyze protocols such as superstrong random
access encodings and the violation of `information causality'.
However, we also show that in such a world \emph{no} entropic
measure can exhibit all properties we commonly accept in a quantum
setting. For example, there exists \emph{no} `reasonable' measure of
conditional entropy that is subadditive.
Finally, we prove a coding theorem for some theories that is
analogous to the quantum and classical setting, providing us with an
appealing operational interpretation.
\end{abstract}
\maketitle

\section{Introduction}

Understanding information in classical and quantum physics has helped us
shed light on the fundamental nature of these theories.
Indeed, it has even been suggested that quantum theory could be more naturally formulated in
terms of its information-theoretic properties~\cite{fuchs, CBH,
Brassard, hardy}. Yet, we have barely scratched the surface of understanding the role
of information in the natural world.
To gain a deeper understanding of information in physical systems, and to help explain \emph{why}
nature is quantum, it is sometimes
instructive to take a step back and view quantum mechanics in a much broader context
of possible physical theories. Many examples are known that indicate that if our world
were only slightly different, our ability to perform information processing tasks
could change dramatically~\cite{wim:nonlocal, boxCrypto, tony:boxCrypto,
juerg:boxCrypto2, howard:convexCrypto, gs:relaxedUR, infoCausality, haenggi:pa}.

However, before we can hope to really investigate general theories
from the perspective of information processing, we first need to
find a way to quantify information. In a quantum and classical
world, this can be done using the von Neumann and Shannon entropy
respectively, which capture our notions of information and
uncertainty in an intuitive way. These quantities have countless
practical applications, and have played an important role in
understanding the power of such theories with respect to information
processing.

Here, we propose a measure of information that applies to \emph{any}
physical theory~\footnote{Although for simplicity we will restrict
our analysis to objects which are finite in an appropriate sense.}
which admits the minimal notions of finite physical \emph{systems},
their \emph{states}, and the probabilistic outcomes of
\emph{measurements} performed on them. Many such theories have been
suggested, each of which shares some aspects with quantum theory,
yet have important differences. For example, we might consider
quantum mechanics itself with a limited set of allowed measurements,
quantum mechanics in a real Hilbert space, generalized probabilistic
theories~\cite{barrett, leifer}, general $C^*$-algebraic theories
\cite{CBH}, box world~\cite{boxMeasurements} (a theory admitting all non-signalling correlations
\cite{PR, boxes}, previously called Generalized Non-Signalling
Theory \cite{barrett}), classical
theories with an epistemic restriction~\cite{spekkens} or theories
derived by relaxing uncertainty relations~\cite{gs:relaxedUR}.

\subsection{A measure of information}

\subsubsection{Entropy}
We propose an entropic measure of information $\ph$ that can be used
in any such theory in Section~\ref{sec:entropyDef}. We will show
that our measure reduces to the von Neumann and Shannon entropy in
the quantum and classical setting respectively. In addition, we show
that it shares many of their appealing intuitive properties. For
example, we show that the quantity is always positive and bounded
for the finite systems we consider. This provides us with a notion
that each system has some maximum amount of information that it can
contain. Furthermore, we might expect that mixing increases entropy.
I.e. that the entropy of a probabilistic mixture of states cannot be
less than the average entropy of its components. This is indeed the
case for our entropic quantity. Another property that is desirable
of a useful measure of information is that it should take on a
similar value for states which are 'close', in the sense that there
exists no way to tell them apart very well. This is the case for the
von Neumann and Shannon entropy, and also for our general entropic
quantity, given one extra minor assumption. Finally, when
considering two different systems $A$ and $B$, one may consider how
the entropy of the joint system $AB$ relates to the entropy of the
individual systems. It is intuitive that our uncertainty about the
entire system $AB$ should not exceed the sum of our uncertainties
about $A$ and $B$ individually. This property is known as
subadditivity and is obeyed by our measure of entropy given one
additional reasonable assumption on the physical theory. Our
entropic quantity thus behaves in very intuitive ways. Yet, we will
see that there exist physical theories for which it is not strongly
subadditive, unlike in quantum mechanics.

Of course, there are multiple ways to quantify information and we
discuss our choice by examining some alternatives and possible
extensions such as notions of accessible information, relative entropy
as well as R{\'e}nyi entropic quantities in Sections~\ref{sec:renyi} and ~\ref{sec:decomp}.

\subsubsection{Conditional entropy and mutual information}

Clearly, it is also desirable to capture our uncertainty about some
system $A$ \emph{conditioned} on the fact that we have access to
another system $B$. This is captured by the \emph{conditional}
entropy, for which we provide two definitions in
Section~\ref{sec:conditionalDef} which are both interesting and
useful in their own right. Based on such definitions we also define
notions of mutual information which allow us to quantify the amount
of information that two systems hold about each other. Our first
definition of conditional entropy is analogous to the quantum
setting, and indeed reduces to the conditional von Neumann entropy
in a quantum world. This is an appealing feature, and opens the
possibility of interesting operational interpretations of this
quantity as in a quantum
setting~\cite{andreas:conditionalInterpretNat,andreas:conditionalInterpretLong}.
Yet, we will see that there exists a theory (called box world) for
which not only the subadditivity of the conditional entropy is
violated, but also where conditioning \emph{increases} entropy.
Intuitively, we would not expect to grow more uncertain when given
additional information, which we could always choose to ignore.

We will hence also introduce a second definition of conditional
entropy, which does not reduce to the von Neumann entropy in the
quantum world. However, it has the advantage that in \emph{any}
theory conditioning \emph{reduces} our uncertainty, as we would
intuitively expect when taking an operational viewpoint.
Nevertheless, even our second definition of the conditional entropy
violates subadditivity.

\subsubsection{Possible properties of the conditional entropy}

Naturally, one might ask whether the fact that both our definitions
of the conditional entropy violate subadditivity is simply a
shortcoming of our definitions.  In Section~\ref{sec:general} we
therefore examine what properties any `reasonable' measure of
conditional entropy can have in principle. By reasonable here we
mean that if given access to a system $B$ we have no uncertainty
about some classical information $A$, then the quantity is '0', and
otherwise it is positive (or even non-zero). We show that under this
simple assumption there exists \emph{no} measure of conditional
entropy in box world that is subadditive or obeys a chain rule.

\subsection{Examples}

To give some intuition about how our entropies can be used outside
of quantum theory, we examine a very simple example in box world in
Section~\ref{sec:boxEntropy}, which illustrates all the peculiar
properties our entropies can have. This is based on a task in which
Alice must produce an encoding of a string $x$, such that Bob can
retrieve any bit of his choosing with some
probability~\cite{as:lowerBound} (known as a random access
encoding). It is known that superstrong random access codes exist in
box world~\cite{gs:relaxedUR}, leading to a violation of the quantum
bound for such encodings~\cite{nayak:rac}.

A similar game was used in~\cite{infoCausality} to argue that one of
the defining characteristics that sets the quantum world apart from
other possibilities (and particularly box world) is that
communication of $m$ classical bits causes information gain of at
most $m$ bits, a principle called `information causality'. In
Section~\ref{sec:infoCausality}, we examine this statement using our
entropic quantity. We notice that it is the failure of subadditivity
of conditional entropy in box world that leads to a violation of the
inequality quantifying 'information causality' given
in~\cite{infoCausality}. We conclude our examples by discussing the
definition of `information causality' more generally.

\subsection{A coding theorem}

In the classical, as well as the quantum setting, the Shannon and
von Neumann entropies have appealing operational interpretations as
they capture our ability to compress information. In
Section~\ref{sec:codingTheorem}, we show that the quantity
$\ph(\cdot)$ has a similar interpretation for some physical
theories. When defining entropy we have chosen to restrict ourselves
to a minimal set of assumptions, only assuming that a theory would
have some notion of states and measurements. To consider compressing
a state or indeed decoding it again, however, we need to know a
little more about our theory. In particular, we first have to define
a notion of `size' for any compression procedure to make sense.
Second, we need to consider what kind of encoding and decoding
operations we are allowed to perform. Given these ideas, and several
additional assumptions on our physical theory, we prove a simple
coding theorem.

\subsection{Outline}

In Section~\ref{sec:assumptions}, we introduce a framework for
describing states, measurements and transformations in general
physical theories, followed in Section~\ref{sec:examples} by some
examples. In Section~\ref{sec:entropy} we then define our entropic
measures of information that can be applied in any theory. Examples
of of how these entropies can be applied in box world can be found
in Section~\ref{sec:boxEntropy}. In Section~\ref{sec:general} we
examine what properties we can hope to expect from a conditional
entropy in box world. Section~\ref{sec:infoCausality} investigates
the notion of 'information causality' in our framework and finally
we show a coding theorem for many theories in
Section~\ref{sec:codingTheorem}. We conclude with many open
questions in Section~\ref{sec:openQuestions}.

\section{An operational framework for physical theories.}\label{sec:assumptions}
We now present a simple framework, based on minimal operational
notions (such as systems, states, measurements and probabilities),
that encompasses both classical and quantum physics, as well as more
novel possibilities (such as `box world')~\cite{leifer, barrett,
hardy, dariano}. Our approach is similar to that in~\cite{leifer},
however it is slightly more general as it does not assume that all
measurements that are mathematically well-defined are physically
implementable, or that joint systems can be characterised by local
measurements.

\subsection{Single systems and states.}
Firstly, we will assume that there is a notion of discrete physical
\emph{systems}. With each system $A$ we associate a set of allowed
\emph{states} $\states_A$, which may differ for each system. We
furthermore assume that we can prepare arbitrary mixtures of states
(for example by tossing a biased coin, and preparing a state
dependent on the outcome), and therefore take $\states_A$ to be a
convex set, with $s_{\textrm{mix}} = p s_1 + (1-p) s_2$ denoting the
state that is the mixture of $s_1$ with probability $p$ and $s_2$
with probability $1-p$. To characterize when two states are the
same, or close to each other, we first need to introduce the notion
of measurements.

\subsection{Measurements}

Secondly, we thus assume that on each system $A$, we can perform a
certain set of allowed measurements $\cE_A = \{\ef\}$. If the system
$A$ is clear from context, we will omit the subscripts and simply
write $\cE$ and $\states$.

With each measurement $\ef$ we associate a set of outcomes
$\cR_{\ef}$, which for simplicity of exposition we take to be
finite. When a particular measurement is performed on a system, the
probability of each outcome should be determined by its state. We
therefore associate each possible outcome $r \in \mathcal{R}_{\ef}$
with a functional $e_r: \states \rightarrow [0,1]$, such that
$e_r(S)$ is the probability of obtaining outcome $r$ given state
$S$. We refer to such a functional as an \emph{effect}. To ensure
that measurement behaves according to our intuition when applied to
mixed states, we require that $e_r(S_{\textrm{mix}} ) = p\, e_r(S_1)
+ (1-p) e_r(S_2)$. This means that each effect can be taken to be
\emph{linear} \footnote{Strictly speaking, we only need the
functional to act linearly on mixtures, requiring it to be affine.
However, it is helpful to extend it to full linearity to deal with
un-normalised states.}. In order for the probabilities of all
measurement outcomes to sum to one, we also require that
\begin{equation}
\sum_{r \in \cR_{\ef}} e_r = u\ ,
\end{equation}
where $u$ is the \emph{unit effect}, which has the property that
$u(S)=1$ for all $S \in \states$. We can thus characterize a
measurement $\ef$ as a set of outcome/effect pairs~\footnote{Note
that here we do not describe a measurement as a set of effects, as
this rules out the possibility of two or more measurement outcomes
corresponding to the same effect, and makes it harder to discuss
coarse-graining, re-labelling or expectation values.}
\begin{align}
\ef = \{(r,e_r) \mid r \in \cR_{\ef} \mbox{ and } \sum_r e_r = u\}\ .
\end{align}
We write
$\mbf{e}(S)$ for the probability distribution over outcomes when
$\mbf{e}$ is performed on a state $S$. Note that in this general
framework, not all measurements that are mathematically well-defined
need be part of a particular physical theory.

One measurement can be equivalent to, or strictly more informative
than, another. Consider two measurements $\mbf{e}$ (with outcomes
$\mathcal{R}_e$ and effects $e_r$) and $\mbf{f}$ (with outcomes
$\mathcal{R}_f$ and effects $f_r$), for which there exists a map
$M:\mathcal{R}_e \rightarrow \mathcal{R}_f$ such that
\begin{equation}
\sum_{\{r \, : \,M(r)=r'\}} e_{r} = f_{r'}   \qquad \forall \; r'\in
\mathcal{R}_f.
\end{equation}
If $M$ is one-to-one it corresponds to a \emph{re-labelling} of the
outcomes. Otherwise, we say that $\mbf{f}$ is a
\emph{coarse-graining} of $\mbf{e}$ (or alternatively that $\mbf{e}$
is a \emph{refinement} of $\mbf{f}$). Because we can always re-label
the outcomes of an experiment according to any map $M$, we assume
that $\mathcal{E}$ is closed under re-labelling and coarse-graining.
This implies that $\mathcal{E}$ always contains the trivial
measurement $\mbf{u}$ (with one outcome corresponding to effect
$u$).

A refinement/coarse-graining is trivial if
\begin{align}\label{eq:prop}
e_r \propto f_{M(r)} \qquad \forall \; r \in \mathcal{R}_e.
\end{align}
In this case, the measurement of $\mbf{e}$ is equivalent to
performing $\mbf{f}$ and obtaining $r'$, then outputting a randomly
selected $r$ satisfying $M(r)=r'$ (where the distribution depends on the proportionality constant in~\eqref{eq:prop}).
Hence the two measurements are
equally informative about the state. In contrast, when $\mbf{e}$ is
a non-trivial refinement of $\mbf{f}$ it offers strictly more
information about the state, and in this case we write $\mbf{e}
\succ \mbf{f}$.
A subset of measurements of particular importance
are the \emph{fine-grained} measurements $\mathcal{E}^* \subseteq
\mathcal{E}$, which have no non-trivial refinements, and are
therefore optimal for gathering information about the state.
Formally,
\begin{equation}
\mbf{e} \in \mathcal{E}^* \; \Leftrightarrow \; \nexists \; \mbf{f}
\in \mathcal{E} \;  : \mbf{f} \succ \mbf{e}
\end{equation}
We will also call an effect $e$ fine-grained if it is part of a fine-grained measurement.
We assume that $\mathcal{E}^*$ is non-empty (i.e. that there exists
at least one finite outcome fine-grained measurement). In quantum
and classical theory this restricts us to the finite-dimensional
case.

\subsection{Transformations}\label{sec:trans}
 As well as preparing states and performing
measurements, it may be possible to perform transformations on a
system. As in the case of effects, in order to behave reasonably
when applied to mixed states, a transformation must correspond to a
linear map $T: \states_A \rightarrow \states_{A'}$ taking allowed
states to allowed states (although the input and output systems may
be of a different type). For each type of system, there will be some
set of allowed transformations $\mathcal{T}$.

We assume that the identity transformation $I$ is allowed, and that
the composition of two allowed transformations is allowed (as long
as the system output by the first transformation is of the same type
as the input to the second). Furthermore, it must be the case that
any allowed transformation followed by an allowed measurement is an
allowed measurement.

We can also combine the notion of transformation with that of
measurement in a natural way to represent non-destructive
measurements~\cite{dariano, barrett}. To incorporate non-destructive
measurements, define the sub-normalised states $\tilde{\states} =
\{p S | 0 \leq p \leq 1, S \in \states\}$. A measurement can then be
described by assigning a subnormalised transformation $t_r: \states
\rightarrow \tilde{\states'}$ to each outcome $r$. Result $r$ occurs
with probability $p_r = u(t_r(s))$ and the post measurement state is
$s_r =t_r(s)/p_r$. However,  we will not need such constructions in
the main part of this paper.

\subsection{Relations between states}

Having introduced measurements, we can now define what it means for
two states to be equal. Given that we are taking an operational
viewpoint, we adopt the intuitive notion that two states $S_1, S_2 \in
\states$ are equal, if and only if there exists no measurement that
distinguishes them. That is,
\begin{equation}
\forall S_1,S_2  \in \states \quad S_1 = S_2 \; \Leftrightarrow
\; \forall \; \mbf{e} \in \mathcal{E} \,:\, \mbf{e}(S_1) =
\mbf{e}(S_2)
\end{equation}

We can also define a natural measure of distance for states $S_0,
S_1 \in \cS$ that directly relates to the probability that we can
distinguish these states using measurements available in our theory,
in analogy to the quantum setting~\cite{cmpNorms}. Suppose we are
given either $S_0$ or $S_1$ with equal probability, and perform a
measurement $\ef$ to distinguish the two cases. Note that the above
implies that any theory that admits at least two possible states has
at least one measurement $\ef$ with two possible outcomes.
Furthermore any such theory must have a measurement $\ef$ with
exactly two outcomes since any theory admits arbitrary coarse-grainings of measurements. We will base our decision on the maximum
likelihood rule, that is, when we obtain outcome $r$, we will
conclude we received state $S_0$ if $e_r(S_0)
> e_r(S_1)$ and $S_1$ otherwise. The probability of distinguishing
the two states using measurement $\ef$ is then given by
\begin{align}
p_{\rm succ}^{\ef} = \frac{1}{2} + \frac{\cdist(\ef(S_0),\ef(S_1))}{2}\ ,
\end{align}
where $\cdist(\ef(S_0),\ef(S_1)) =
\frac{1}{2}\sum_{r \in \cR_e}
|e_r(S_0) - e_r(S_1)|$ is the classical statistical distance between
the probability distributions $\ef(S_0)$ and $\ef(S_1)$. We now
define the distance as
\begin{align}\label{eq:distanceMeasure}
\dist(S_0, S_1) := \sup_{\ef} \cdist(\ef(S_0),\ef(S_1))\ .
\end{align}

By the above, we see that this measure of distance has an appealing
operational interpretation because it directly captures our ability
to distinguish the two states $S_0$ and $S_1$ using any available
measurement (see appendix \ref{app:distance},
Lemma~\ref{lem:distanceMeasure} for details). In the quantum
setting, it thus directly reduces to the well-known trace distance.

\subsection{Multi-partite systems} Suppose that we have two systems
$A$ and $B$, each of which may admit different sets of states and measurements.
We allow that two individual systems can be combined into a \emph{composite} system $AB$,
which we can treat
as a new type of
system having its own set of allowed states, measurements, and
transformations just as in the single-system case. However, these
sets must bear some relation to those of the component subsystems.

With respect to states, we would like it to be possible to
independently prepare any state $S_A \in \mathcal{S}_A$ of system A
and $S_B \in \mathcal{S}_B$ of system B. This corresponds to a
\emph{product state} of the composite system, which we denote by
$S_{AB} = S_A \otimes S_B \in \mathcal{S}_{AB}$. Note that at this
point we have not proved that $\otimes$ corresponds to a tensor
product in the usual sense \footnote{It is possible to prove this
with the additional assumption that multi-partite systems are
completely characterised by product measurements~\cite{barrett,
leifer}. However, this rules out some potentially reasonable
theories, such as quantum theory in a real Hilbert space, and we
will not need to make this additional assumption here.}, but we
would nevertheless expect that it is distributive for mixtures and
associative. We make use of the standard terminology that states are
\emph{separable} if they can be written as a mixture of product
states, and \emph{entangled} otherwise. To avoid excessive
subscripts when dealing with multiple systems, we will usually refer
to the state of systems $AB$ and $B$ directly by these letters,
rather than the more cumbersome $S_{AB}$ and $S_{A}$ (e.g.
$\mbf{e}(S_{AB}) = \mbf{e}(AB)$ etc. ).

Similarly, we would expect to be able to perform a measurement
$\mbf{e} \in \mathcal{E}_A$ and $\mbf{f} \in \mathcal{E}_B$, giving
a product measurement which we denote by $\mbf{g} = \mbf{e} \otimes
\mbf{f} \in \mathcal{E}_{AB}$ (with outcome set $\mathcal{R}_{\gef} = \mathcal{R}_{\ef} \times
\mathcal{R}_{\fef}$ and effects $g_{ij} = e_i \otimes f_j$). By considering
coarse-graining and tri-partite systems, we would again expect
$\otimes$  to be distributive and associative. When applying a
product measurement to a product state we furthermore require that
\begin{align}\label{eq:productOutcomes}
(e_i \otimes f_j)(A \otimes B) = e_i(A)f_j(B)\ .
\end{align}

When considering multiple systems, we can consider what happens if we
only measure some of these systems. Note that this means that we perform
a measurement consisting of a unit effect on some of these systems.
This only makes sense if marginal states are well defined and we hence
assume that
even when a bipartite state is entangled each part
is an allowed marginal state. We can thus have
\begin{equation}
\forall\, (AB) \in \mathcal{S_{AB}}, \exists A \in \mathcal{S}_A :
\forall\, \mbf{e} \in \mathcal{E}_A, \mbf{e}(A) = (\mbf{e}\otimes
\mbf{u})(AB).
\end{equation}
Furthermore, in the case in which B performs a measurement on his
subsystem and obtains result $r$ (corresponding to an effect $e_r$)
we would expect A's subsystem to `collapse' to an allowed state
$A_{|r} \in \mathcal{S}_A$. We will denote such a state as
\begin{equation}
A_{|r} = \frac{(I \otimes e_r) (AB)}{e_r(B)}.
\end{equation}

Finally, a crucial constraint on multi-partite systems is the
existence of product transformations $T_A \otimes T_B \in
\mathcal{T}_{AB}$. In a variant of quantum theory in which all
positive (rather than completely positive) trace-preserving maps are
allowed transformations, this would prevent the existence of
entangled states.

\section{Example theories} \label{sec:examples}

In this section we show how quantum theory and classical probability
theory fit into the framework defined above, and also describe the theory known as `box world' \cite{barrett, boxMeasurements}, which admits
all non-signalling correlations \cite{boxes, PR}, and was one of the main
motivations for this work.

\subsection{Classical probability Theory} In classical probability
theory, a state $S$ corresponds to a probability distribution $p_i$
over a finite set of elements. The effects correspond to linear functionals
of the form
\begin{equation}
e_r(S) = \sum_i q_r^i p_i
\end{equation}
for any $q_r^i \in [0,1]$. Note that the unit effect corresponds to
$q^i=1 \; \forall \; i$. Normalisation of measurements therefore
requires $\sum_r q_r^i=1 \; \forall \; i.$ Transformations
correspond to stochastic maps.

\subsection{Quantum Theory} In quantum theory, the convex set of states
are the density operators $S= \rho$ (trace-1 positive operators),
and effects correspond to linear functionals of the form
\begin{equation}
e_r(S) = \tr( \rho E_r )
\end{equation}
where $E_r$ is a positive operator. All measurements satisfying the
normalisation constraint
\begin{equation}
\sum_r e_r = u \; \Rightarrow \; \sum_r E_r = I
\end{equation}
are allowed, and the fine-grained measurements are those for which
all $E_r$ are rank 1 operators. The allowed transformations
represent completely positive trace-preserving maps~\cite{nielsen&chuang:qc}.

\subsection{Restricted Quantum/classical theories}
Note that unlike other approaches~\cite{leifer, barrett} our
framework also encompasses real Hilbert space quantum mechanics.
Furthermore, because we do not assume that all well-defined
operations are physically realizable, it can be used to study
quantum or classical theory with a restricted set of states,
measurements and transformations (for an interesting example in the
classical case consider Spekkens' toy model~\cite{spekkens}). The
entropies we would assign in such cases would differ from the
standard von Neumann entropy, and may be interesting to study.

\subsection{Box world} In box world, the state of a single system $X$
corresponds to a conditional probability distribution
$S=P(\xout|\xin)$ where $\xin$ and $\xout$ are elements of a finite
set of `inputs' and `outputs' respectively. The intuition is that
there is a special set of measurements on each system represented
by $\xin$ (referred to as \emph{fiducial} measurements), and that
any probability distribution for these measurements corresponds to
an allowed state. We represent a system $X$ with $k$ possible inputs
$\xin$ and $m$ possible outputs $\xout$ by
\begin{center}
\includegraphics{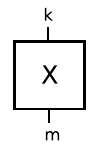}
\end{center}

In the special case in which there is only one possible input, the
conditional probability distribution reduces to the standard
unconditional probability distribution $P(\xout)$, and we omit the
input line to the box in the diagram. Thus box world contains
classical probability theory as a special case, and we will use such
\emph{classical boxes} to represent classical information in our
treatment of information-theoretic protocols in box world.

A multi-partite state in box world corresponds to a joint
conditional probability distribution $P(\xout^1 \xout^2 \ldots
\xout^N | \xin^1 \xin^2 \ldots \xin^N)$ with a separate input and
output for each system. Aside from the usual constraints of
normalisation and positivity, the allowed states must also satisfy
the non-signalling conditions: That the marginal probability
distribution obtained by summing over $\xout^k$,
\begin{align}
\sum_{\xout^k} P(\xout^1 \ldots \xout^k \ldots \xout^N | \xin^1
\ldots \xin^k \ldots \xin^N),
\end{align}
is independent of $\xin^k$ for all $k$. This means that the other
parties cannot learn anything about a distant party's measurement
choice from their own measurement results. A bipartite state of
particular interest is the PR-box state~\cite{PR,PR2,PR3}, for which
all inputs and outputs are binary, and the probability distribution
is
\begin{equation} \label{eq:PR}
P_{PR} (\xout^1 \xout^2 | \xin^1 \xin^2) = \left\{
\begin{array}{ccl} \frac{1}{2} & : &
\xout^1 \oplus \xout^2 = \xin^1 \cdot \xin^2 \\
 0& :  & \mathrm{otherwise}
\end{array} \right.
\end{equation}
where $\oplus$ denotes addition modulo 2. This state is `more
entangled' than any quantum state, yielding correlations that
achieve the maximum possible value of 4 for the
Clauser-Horne-Shimony-Holt (CHSH) expression~\cite{CHSH}, compared
to $ \leq2\sqrt{2}$ for quantum theory (Tsirelson's bound
\cite{tsirel:original}), and $\leq 2$ for classical probability
theory. We represent entanglement between systems in box world by a
zigzag line between them, and classical correlations (i.e. separable
but non-product states) by a dotted line.
\begin{center}
\includegraphics{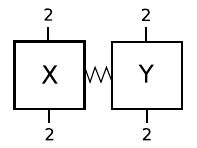}
\includegraphics{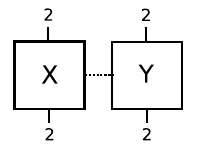}
\end{center}

In box world, we allow all mathematically well-defined measurements
and transformations to be physically implemented. Writing
$\vxout=(\xout^1, \xout^2, \ldots, \xout^N)$ and $\vxin=(\xin^1, \xin^2, \ldots,
\xin^N)$, all effects take the form
\begin{equation}
e_r(S) = \sum_{\vxout,\vxin} Q_r(\vxout|\vxin)
P(\vxout|\vxin),
\end{equation}
where $Q_r(\xout|\vxin)$ can be taken to be positive~\cite{barrett}.
The effect $e^{\vxin'}_{\vxout'}$ corresponding
to performing joint fiducial measurements $\vxin'$ and obtaining
results $\vxout'$ is represented by $Q_{\vxout'}(\vxout|\vxin)
= \delta_{\xin \xin'} \delta_{\xout,\xout'}$. Because of the positivity of $Q_r$,
any effect can be expressed as a weighted sum of such fiducial
measurement effects. It follows that a measurement is fine-grained
if and only if each of its effects is proportional to some
$e^{\vxin}_{\vxout}$, and that products of fine-grained
measurements are themselves fine-grained.

\section{Generalized entropies}\label{sec:entropy}

The Shannon entropy $H(\vec{p})=-\sum_i p_i \log p_i$ and von
Neumann entropy $S(\rho) = - \tr(\rho \log \rho)$ are extremely
useful tools for analyzing information processing in a classical or
quantum world. Here, we would like to define an analogous entropy
for general probabilistic theories which reduces to $H(\vec{p})$ and
$S(\rho)$ for classical probability theory and quantum theory
respectively. We would also like our new entropy to retain as many
of the mathematical properties of the Shannon and von Neumann
entropy as possible. Not only will this help our new entropy conform
to our intuitive notions, but it will make it easier to prove
general results using these quantities, and transfer known results
to the general case. Note that although we can use any base for the
logarithm in the definition of the Shannon and von Neumann entropies
(as long as we are consistent), in what follows we will use base 2
(i.e. $\log = \log_2$) throughout.

\subsection{Entropy}\label{sec:entropyDef}
We now give a concrete definition of entropy for any physical
theory, which satisfies the above desiderata. Other definitions are
certainly possible, and we will consider one alternative (based on
mixed state decomposition) in Section \ref{sec:decomp}.
However, the following definition has many appealing properties.

Given any state $S \in \cS$, we define its entropy $\ph(S)$ by
\begin{align}\label{eq:entropyDef1}
\ph(S) := \inf_{\ef \in \mathcal{E}^*} \sh(\ef(S))\ ,
\end{align}
where the infimum is taken over all fine-grained measurements $\ef
\in \mathcal{E}^*$ on the state space $\cS$ and $\sh(\ef(S)) = -
\sum_{r\in \cR_{\ef}} e_r(S) \log e_r(S)$ is the Shannon entropy of
the probability distribution $\ef(S)$ over possible outcomes of
$\ef$. This has an intuitive operational meaning as the minimal
output uncertainty of any fine-grained measurement on the system.
Note that for information-gathering purposes, the best measurements
are always fine-grained, and without restricting to this subset the
unit measurement would always be optimal (giving zero outcome
uncertainty). Furthermore note that trivial refinements of $\mbf{e}$
always generate a higher output entropy, so it is sufficient to only
consider measurements in the infimum that have no parallel effects.

In appendix \ref{app:entropy}, we prove that $\ph$ retains
several important properties of the Shannon and von Neumann entropy.
In particular, we show:
\begin{enumerate}
\item (\emph{Reduction}) $\ph$ reduces to the Shannon entropy for classical probability theory, and
the von Neumann entropy for quantum theory.
\item (\emph{Positivity and boundedness}) Suppose that the minimal number of outcomes for a fine-grained measurement in $\mathcal{E}_S^*$ is $d$.
Then for all states $S \in \cS$,
\begin{equation} \label{eq:bounded}
 \log(d) \geq \ph(S) \geq 0.
\end{equation}
\item (\emph{Concavity})
For any $S_1, S_2 \in \cS$ and any mixed state $S_{\textrm{mix}} = p S_1 +
(1-p) S_2 \in \cS$:
\begin{equation}
\ph(S_{\textrm{mix}}) \geq p \ph(S_1) + (1-p) \ph(S_2)\ .
\end{equation}
\item (\emph{Limited Subadditivity})
Consider a theory with the additional property that fine-grained
measurements remain fine-grained for composite systems. i.e.
\begin{equation} \label{eq:fineRefinement}
\mbf{e} \in \mathcal{E}_A^*, \mbf{f} \in \mathcal{E}_B^* \;
\Rightarrow \mbf{e} \otimes \mbf{f} \in \mathcal{E}_{AB}^*.
\end{equation}
This is true in quantum theory, classical theory, and box world.
When \eqref{eq:fineRefinement} holds, then for any bipartite state
$AB \in \mathcal{S}_{AB}$ and reduced states $A \in \cS_A$ and $B
\in \cS_B$
\begin{equation}\label{eq:subadd}
\ph(A) + \ph(B) \geq \ph(AB)\
\end{equation}
\item (\emph{Limited Continuity}).
Consider a system for which all allowed measurements have at most
$D$ outcomes, or for which restricting the allowed measurements to
have at most $D$ outcomes does not change the entropy of any state.
This is true in quantum theory, with $D=d=\dim(\mathcal{H})$, and
also in box world and classical theory. Then we can prove an
analogue of the Fannes inequality~\cite{fannes:inequ, hayashi:book},
which says that the entropy of two states which are close does not
differ by too much. In particular, given $S_1,S_2 \in \cS$
satisfying $\dist(S_1, S_2) < 1/e$,
\begin{align}
|\ph(S_1) - \ph(S_2)| \leq \dist(S_1, S_2) \log \left(\frac{D}{\dist(S_1,
S_2)}\right).
\end{align}
\end{enumerate}
We will also see in section~\ref{sec:codingTheorem} that $\ph$ has
an appealing operational interpretation as a measure of
compressibility for some theories.

However, one property of the von Neumann entropy that does not carry
over to $\ph$ is strong subadditivity~\cite{nielsen&chuang:qc}. In particular, we
will see in section \ref{sec:boxEntropy} there exists a tripartite
state in box world such that
\begin{align}
\ph(ABC) + \ph(C) > \ph(AC) + \ph(BC)
\end{align}
\subsection{Conditional entropy and mutual information}\label{sec:conditionalDef}

\subsubsection{A standard definition}

Based on the entropy $\ph$, we can also define a notion of
conditional entropy. In analogy to the von Neumann
entropy~\cite{adami:negative}, we define the conditional entropy of
a general bipartite state $AB \in \mathcal{S}_{AB}$ with reduced
states $A \in \cS_A$ and $B \in \cS_B$ by
\begin{align}\label{eq:conditionalEntropy1}
\ph(A|B) := \ph(AB) - \ph(B)\ .
\end{align}
This has the nice property that for quantum or classical systems it
reduces to the conditional von Neumann and Shannon entropies
respectively. In some theories (including quantum theory but not
classical probability theory), $\ph(A|B)$ can be negative, which is
strange, but opens the way for an appealing operational
interpretation as in the quantum setting~\cite{andreas:conditionalInterpretNat}.

However, unlike in quantum theory, we will see that
$\ph(\cdot|\cdot)$ has the
counterintuitive property that it can \emph{decrease} when
`forgetting' information in some probabilistic theories. In
particular, the violation of strong subadditivity for $\ph$ in box
world implies that it is possible to obtain $\ph(A|BC)> \ph(A|B)$,
and that $\ph(\cdot|\cdot)$ is not subadditive. These properties
will motivate us to consider an alternative definition of the
conditional entropy below. However, we will
show that no `reasonable' entropy in box world can have all the
appealing properties of the conditional von Neumann entropy.

In analogy to the quantum case, we can also define the \emph{mutual information} via
\begin{eqnarray}
\ih(A;B) &:=& \ph(A) +\ph(B)- \ph(AB).\\
&=& \ph(A)- \ph(A|B) = \ph(B) - \ph(B|A) \nonumber
\end{eqnarray}
This quantity will be positive whenever subadditivity holds, and
reduces to the usual mutual information in the quantum and classical
case. Similarly, we may define a notion of \emph{accessible
information} analogous to the quantum setting as
\begin{eqnarray}
\ih_{\rm acc}(A;B) &:=& \sup_{\ef \in \cE_A, \fef \in \cE_B} \ii(\ef(A);\fef(B))\ ,
\end{eqnarray}
where $\ii$ is the classical mutual information.

\subsubsection{An alternative definition}

Given the problems observed with the previous definition in some theories, we
now define a second form of conditional entropy based on $\ph$, which sometimes
captures our intuitive notions about information in a nicer way. For
any bipartite state $AB \in \cS_{AB}$ with reduced states $A \in
\cS_A$ and $B \in \cS_B$ we define
\begin{align}\label{eq:conditionalEntropy2}
\pht(A|B) := \inf_{\fef \in \mathcal{E}_B} \sum_j f_j(B)
\ph(A_{|j})\,
\end{align}
where the infimum is taken over all measurements on $B$, and
$A_{|j}$ is the reduced state of the first system conditioned on
obtaining measurement outcome $j$ when performing $\fef$ on the
second system. This definition has the appealing property that
conditioning on more systems always reduces the entropy, that is,
$\ph(A) \geq \pht(A|B) \geq \pht(A|BC)$ (see appendix
\ref{app:condEntropy}, Lemma~\ref{lem:conditioningReducesEntropy}),
and it reduces to the conditional Shannon entropy in the classical
case. Note, however, that $\pht(\cdot|\cdot)$ does not reduce to the
conditional von Neumann entropy in the quantum setting, as it is
always positive. Furthermore, we will see in
section~\ref{sec:general} that it is not subadditive, and does not
obey the usual chain rule. (even though a limited form of chain rule holds
in box world as we show in the appendix Section~\ref{sec:boxChain}).
Nevertheless $\pht(\cdot|\cdot)$ seems
quite a natural entropic quantity, and its corresponding quantum version
has found an interesting application in the study of quantum correlations~\cite{discord}.

We can also define a corresponding information quantity via
\begin{align}
\iht(A;B) = \ph(A) - \pht(A|B).
\end{align}
which is always positive. However, unlike $\ih(A;B)$, this
definition is not symmetric and hence it cannot really be considered
`mutual information'. Instead, $\iht(A;B)$ captures the amount of
information that $B$ holds about $A$.

\subsection{Other entropic quantities}\label{sec:renyi}

For cryptographic purposes, such as in the setting of device
independent security for quantum key distribution, it is useful to
define the following R{\'e}nyi entropic variants of $\ph$. More
precisely, we define
\begin{align}
\ph_\alpha(S) := \inf_{\ef \in \mathcal{E}^*} \sh_\alpha(\ef(S))\ ,
\end{align}
where
$\sh_\alpha(\ef(S)) = \frac{1}{1-\alpha} \log\left(\sum_j (\ef(S)_j)^\alpha\right)$ is the R{\'e}nyi entropy of order $\alpha$.
Note that $\sh_1(S) = \sh(S)$ (taking the limit of $\alpha \rightarrow 1$).
These quantities can also be useful in order to bound the value of $\ph(\cdot)$ itself as
for any state $S \in \cS$ and $\alpha < \beta$ we have $\ph_\beta(S) \geq \ph_\alpha(S)$.

To define a notion of relative entropy, we adopt a purely
operational viewpoint. Suppose we are given $N$ copies of a state
$S_1$ or a state $S_2$, and let
\begin{align*}
S^N_1 &:= S_1^{\otimes N}\\
S^N_2 &:= S_2^{\otimes N}\ .
\end{align*}
Classically, as well as quantumly, the relative entropy captures our
ability to distinguish $S^N_1$ from $S^N_2$ for large $N$. Note that
to distinguish the two cases, it is sufficient to coarse grain any
measurement to a two outcome measurement $\ef=\{(1,e_1), (2,e_2)\}$,
where without loss of generality we associate the outcome `1' with
the state $S^N_1$ and `2' with $S^N_2$. Then $e_1(S^N_2)$ denotes
the probability that we conclude that the state was $S^N_2$, when
really we were given $S^N_1$. Similarly, $e_2(S^N_1)$ denotes the
probability that we falsely conclude that the state was $S^N_2$. In
what is called asymmetric hypothesis testing, we wish to minimize
the error $e_1(S^N_2)$ while simultaneously demanding that
$e_2(S^N_1)$ is bounded from above by a parameter $\eps$. Here we
fix $\eps = 1/2$. We therefore want to determine
\begin{align}
p_N := \inf_{\ef} \{e_1(S^N_2) | e_2(S^N_1) \leq 1/2\}
\end{align}
In a quantum setting, it has been shown that
the quantum relative entropy is directly related to this quantity
via the quantum Stein's lemma~\cite{hpetz,ogawa:stein,brandao:stein}, which states that
we have
\begin{align} \label{eq:rel_ent}
D(S_1||S_2) = \lim_{N \rightarrow \infty} - \frac{\log p_N}{N}\ .
\end{align}
This is a deep result giving a clear operational interpretation to
the relative entropy, telling us that in the large $N$ limit the
probability of making the error $p_N$ decreases exponentially with
$D(S_1||S_2)$. Furthermore, as it is expressed in operational
terms, we can simply adopt  (\ref{eq:rel_ent}) as our definition of relative
entropy in any theory for which the limit is well defined. Thus we recover the
usual value in the quantum (and classical) case, and in all other
theories we still capture the same operational interpretation.

Note also that our choice of $\eps = 1/2$ was quite arbitrary, and one may consider
a family of relative entropies, one for each choice of $\eps$. In quantum theory, these are all
equivalent~\cite{brandao:stein}, but they may yield different values in other theories.

\subsection{Decomposition entropy}\label{sec:decomp}

Although the entropy $\ph$ has several appealing properties, and
seems quite intuitive, it is nevertheless interesting to consider
alternative notions of entropy for general theories. One seemingly
natural alternative is the decomposition entropy, which measures the
mixedness of a state.

There is a special subset of states $\mathcal{S}^* \subseteq
\mathcal{S}$ which cannot be obtained by mixing other states:
\begin{align}
S \in \mathcal{S}^*  \; \Leftrightarrow \; \nexists \, S_1, S_2 \in
\mathcal{S}, p\in(0,1) : S=p S_1 + (1-p) S_2.
\end{align}
$\mathcal{S}^*$ form the extreme points of $\mathcal{S}$ and are
referred to as \emph{pure states} (with the remaining states being
\emph{mixed}). Suppose that any state in $\mathcal{S}$ can be
decomposed into a finite sum of pure states. Then we can define the
entropy of a state by the minimal Shannon entropy of its
decompositions into pure states. Define a decomposition $\mbf{D}(S)$
of a state $S \in \cS$ as a probability distribution over the set of
pure states that is non-zero for only a finite set of states $S_i
\in \mathcal{S}^*$ with probabilities $p_i \in (0,1]$ such that
$\sum p_i S_i = S$. Then define the decomposition entropy as
\begin{equation}
\bh(S) := \inf_{\mbf{D}(S)} \sh(\mbf{D}(S)).
\end{equation}

Like our previous entropy definition, we show in appendix
\ref{app:decomposition} that $\bh$ reduces to the Shannon and von
Neumann entropy in classical probability theory and quantum theory
respectively. However, it has a number of unappealing properties
when compared with $\ph$. In particular it is neither concave nor
subadditive, as revealed by explicit counterexamples from box world
given in appendix \ref{app:decomposition}.

After studying simple examples in box world, it seems that $\bh$ is
a less intuitive and helpful measure of uncertainty than $\ph$. For
this reason, although $\bh$ may play an important role in
discussions of entanglement or purity in many generalized theories,
and may also lead to interesting operational interpretations, we do
not discuss it further here.

\section{Examples in Box World}\label{sec:boxEntropy}

We now investigate how our entropic quantity $\ph(\cdot)$ behaves in
box world with a simple, yet illustrative, example.

To first gain some intuition on how $\ph$ behaves in such a setting,
consider a trivial classical system $X$ which admits only one
possible measurement and outputs $2$ possible values $\xout \in \01$
each which probability $1/2$.
\begin{center}
\includegraphics{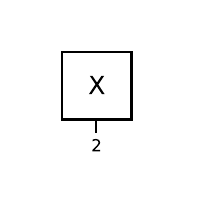}
\end{center}
Clearly, since the system admits only one possible measurement $\ef$, we have
\begin{align}
\ph(X) = \sh(\ef(X)) = \sh((1/2,1/2)) = 1\ .
\end{align}
Consider now a PR-box (a bipartite system in the state
(\ref{eq:PR}))
\begin{center}
\includegraphics{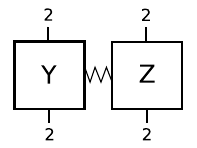}
\end{center}
where Alice holds system $Y$ (with binary input $\yin$ and output $\yout$)
and Bob holds system $Z$ (with binary input $\zin$ and output $\zout$).
Note that the fine-grained measurements on the entire system
correspond to a sequence of fiducial measurements on the two
subsystems (where the choice of input to the second subsystem may
depend on the output of the first)\cite{barrett}, and the outcome is
the output of both measurements. The minimal entropy for the joint
system can be obtained by inputting `0' into both boxes, giving
outputs `00' or `11' each with probability $1/2$ (in fact, any other
fine-grained measurement is equally good), and the marginal states
yield a random output bit for any input. Hence we have that
\begin{align}
\ph(Y) = \ph(Z) = \ph(YZ) = 1.
\end{align}
We now consider a scenario for which it is known that PR-boxes yield
an advantage over the quantum setting in terms of information
processing. The basis of our example is a simple non-local game in
which Alice is given a random `parity' bit $x$, and has to output
two bits $x_0$ and $x_1$ satisfying $x_0 \oplus x_1 = x$ (where
$\oplus$ denotes addition modulo 2). Then, without receiving any
communication from Alice, Bob is given a random target bit $t$ and
has to successfully output $x_t$~\cite{as:racNotes}. This game is
equivalent to the CHSH-game~\cite{CHSH, as:lowerBound}.

We begin with Alice having the parity bit (which we model by a
classical box in the state $X$ described above), and Alice and Bob
sharing a PR-box in the state $YZ$.
\begin{center}
\includegraphics{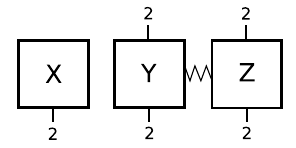}
\end{center}
Now Alice performs the following procedure, which corresponds to an
allowed transformation in box world. She measures the parity bit $X$
to obtain $x := \xout$, then uses this as the input to her part of the
PR-box, setting $\yin = x$ and obtaining outcome $\yout$. Finally, she
prepares two new classical bits $x_0 = \yout$ and $x_1= x \oplus \yout$
(represented by classical boxes $X_0, X_1$). Note that because of
the correlations inherent in the PR box, the output of Bob's system
will now be described by $\zout =\yin \cdot \zin \oplus \yout = (x_0 \oplus x_1) \cdot \zin \oplus
x_0 = x_{\zin}$. Hence the state of $X_0X_1Z$ after this procedure is the
classically correlated state:
\begin{center}
\includegraphics{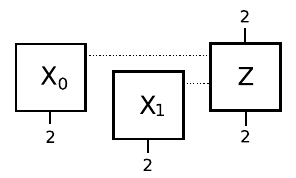}
\end{center}
\begin{equation} \label{eq:bwstate}
P(x_0 x_1 \zout|\zin) = \left\{ \begin{array}{ccl} \frac{1}{4} & : &
\zout = x_{\zin} \\ 0 &:& \textrm{otherwise} \end{array} \right.
\end{equation}
Given any target bit $t$, Bob can win the game by setting $\zin = t$ and
outputting the result $\zout=x_t$. We can think of Bob's system as a
perfect random access encoding of the two-bit string $x_0x_1$
\cite{as:lowerBound,gs:relaxedUR}.

Consider the entropies of the state $X_0X_1Z$. All of the individual
systems yield a random output bit, giving
\begin{align}
\ph(X_0) = \ph(X_1) = \ph(Z) = 1\ ,
\end{align}
and $x_0$ and $x_1$ are independent random bits, so
\begin{align}
\ph(X_0X_1) = 2\ .
\end{align}
Also note that we have
\begin{align}
\ph(X_0 X_1 Z) = 2\ ,
\end{align}
since for any input $\zin$, the output $\zout$ will be perfectly correlated
with one of the other bits (giving only 2 independent random output
bits). Finally, because we can make $\zout$ perfectly correlated with
either of the remaining bits we have
\begin{align}
\ph(X_0 Z) = \ph(X_1 Z) = 1,
\end{align}
where the optimal measurements are $\zin=0$ and $\zin=1$ respectively.

These entropy values all seem very intuitive (Note in contrast that
for the decomposition entropy $\bh(X_0 Z)=2)$. However, they violate
several natural properties of the Shannon and von Neumann entropies.

\para{Strong subadditivity} First of all, it is easy to see from the
above that
\begin{align}
\ph(X_0X_1Z) + \ph(Z) > \ph(X_0Z) + \ph(X_1Z),
\end{align}
which violates strong subadditivity.
We now turn to the two possible forms of conditional entropy that we defined,
where our simple example clearly illustrates their differences.

\subsection{Standard conditional entropy}

First of all, we consider the standard form of conditional entropy,
which reduces to the von Neumann entropy in the quantum settings. By
the above, we can immediately see that it has the following
interesting properties.

\para{Subadditivity of the conditional entropy}
Using \eqref{eq:conditionalEntropy1} we deduce that
\begin{align}
\ph(X_0|Z) = \ph(X_1|Z) =0, \; \ph(X_0X_1|Z)=1
\end{align}
which seems intuitive, as we can perfectly predict the output of
either $X_0$ or $X_1$ (but not both) using $Z$. However, this yields
a violation of subadditivity for the conditional entropy, as
\begin{align}
\ph(X_0X_1|Z) > \ph(X_0|Z) +\ph(X_1|Z).
\end{align}
This may seem rather bizarre at first glance, however, we will see
in Section~\ref{sec:general} that no `reasonable' measure of
conditional entropy in box world is subadditive, unlike the von
Neumann entropy.

It is also interesting to consider the corresponding mutual
information quantities, which are
\begin{align}
\ih(X_0;Z) = \ih(X_1;Z) = \ih(X_0 X_1;Z) =1.
\end{align}
Again, these seem intuitive, as we can extract one bit of
information about either $X_0$ or $X_1$ or the pair $X_0 X_1$ from
$Z$.

It may be tempting to conclude that the point at which $\ph(X_0
X_1|Z)$ becomes subadditive (or equivalently, where $\ph(X_0 X_1 Z)$
becomes strongly subadditive) is exactly when the PR-box is weakened
to obey Tsirelson's bound. Note that our trivial example only shows
that PR-boxes which are more than $\approx 0.89 > 1/2 +
1/(2\sqrt{2})$ correct do not obey subadditivity. However, note that
constraining non-local boxes to obey Tsirelson's bound alone is
insufficient to reduce box world to quantum theory (e.g. each
quantum system admits a continuum of fine-grained measurements
whereas any box admits only a finite set).

\para{Conditioning can increase entropy}
Our small example also emphasizes another curious property
of the conditional entropy. By definition,
\begin{align}
\ph(X_0|X_1 Z) = \ph(X_0 X_1 Z) - \ph(X_1 Z) = 1.
\end{align}
But this is strange, because we can perfectly determine the output
of $X_0$ given $Z$. Furthermore, since $\ph(X_0|Z) = 0$, we then
clearly have
\begin{align}
\ph(X_0|X_1 Z) > \ph(X_0|Z)\ ,
\end{align}
which means that `forgetting information', namely discarding $X_1$,
can \emph{decrease} uncertainty. Again, it may seem that this is a
consequence of not choosing the `correct' definition of entropy.

\subsection{Alternative conditional entropy}

Reevaluating the conditional entropies of the previous section using
this new definition we find that
\begin{align}
\pht(X_0|Z) = \pht(X_1|Z) = 0, \;  \pht(X_0 X_1 |Z) = 1\
\end{align}
as before, hence this new measure still violates subadditivity.
However we now have
\begin{align}
\pht(X_0|Z X_1) = 0\ ,
\end{align}
as we would intuitively expect. This means that conditioning on
$X_1$ no longer increases the entropy. However, it generates a
violation of the chain rule
\begin{align}
\pht(X_0 X_1|Z) \neq \pht(X_1|Z) + \pht(X_0|X_1 Z).
\end{align}
On balance though, this measure of conditional entropy seems more
reasonable than the original one in this example.

\section{Properties of conditional entropies in box world}\label{sec:general}

We now show that \emph{any} `reasonable' measure of the conditional
entropy in box world will necessarily defy our intuition about
information in several ways.

Intuitively, the goal of any entropic quantity is to capture the
degree of uncertainty we have about a system, possibly given access
to some additional information. We assign a label $A$ to the system
of interest and use $B$ to denote any additional systems or
information available to us. For simplicity, let us suppose that $A$
corresponds to some classical information (i.e. it is a state of a
classical box). Let $\h(A|B)$ denote some entropic quantity that
quantifies our uncertainty about $A$ given $B$. If we were able to
determine $A$ with certainty given access to $B$ (i.e. to determine
the precise output of the classical box $A$), we would intuitively
say that there is no uncertainty and the quantity $\h(A|B)$ should
vanish. Conversely, if we cannot determine $A$ given $B$, but will
necessarily have some residual uncertainty, then the quantity
$\h(A|B)$ should be positive. Motivated by this intuition in
quantifying uncertainty we demand the following two properties to
hold for any `reasonable' measure of uncertainty when $A$ is
classical.
\begin{enumerate}
\item[\{1\}] If the output of $A$ can be obtained from $B$ with certainty, $\h(A|B) = 0$.
\item[\{2\}] If the output of $A$ cannot be obtained from $B$ with certainty, then $\h(A|B) > 0$.
\end{enumerate}

In the classical and quantum world, all commonly used entropic
quantities satisfy these conditions (given that $A$ is classical).
In both such worlds, there also exist entropic quantities that are
subadditive and obey a chain rule, for example the conditional
Shannon and von Neumann entropies. In box world, $\pht(A|B)$ is
`reasonable' according to this definition, while $\ph(A|B)$ is
`unreasonable'. Curiously, it turns out that in box world there
cannot be \emph{any} reasonable measure of conditional entropy that
obeys conditions \{1\} and \{2\}, but at the same time is
subadditive or obeys a chain rule.

\para{Subadditivity of the conditional
entropy} Consider the state of the two classical bits $A=X_0 X_1$
and Bob's binary input/output box $B=Z$ described by
(\ref{eq:bwstate}) in the previous section. We now show that in this
case \emph{no} reasonable measure of entropy that obeys properties
\{1\} and \{2\} is subadditive. First of all, note that Bob can
determine one of the bits perfectly, given access to $Z$. Therefore
from condition \{1\}, we obtain that
\begin{align}\label{eq:subaddRHS}
\h(X_0|Z) = \h(X_1|Z) = 0\ .
\end{align}
However, since Bob cannot determine the parity of the two bits, he
certainly cannot learn both bits perfectly and hence from condition
\{2\} we have
\begin{align}\label{eq:subaddLHS}
\h(X_0 X_1|Z) > 0\ .
\end{align}
In order for subadditivity to hold, we would need
that
\begin{align}
\h(X_0 X_1|Z) \leq \h(X_0|Z) + \h(X_1|Z)\ ,
\end{align}
which using~\eqref{eq:subaddRHS} and~\eqref{eq:subaddLHS} leads to a
contradiction. Note that subadditivity could still hold, if the
quantity $\h(X_0 X_1|Z)$ were negative.

\para{Chain rule for the conditional entropy}
We now show that  a chain rule is impossible in box world for any
entropic quantity that satisfies \{1\} and \{2\}. In fact, for the
purposes of this proof it is sufficient to replace \{2\} by the
weaker assumption
\begin{enumerate}
\item[\{2'\}] If the output of $A$ cannot be obtained from $B$ with certainty, then $\h(A|B)\neq 0$.
\end{enumerate}
Note that for the state described by (\ref{eq:bwstate}), condition
\{1\} gives us
\begin{align}\label{eq:removeX1}
\h(X_0|Z,X_1) = \h(X_0|Z) =0
\end{align}
because $x_0$ can be obtained perfectly from $B=Z$ or $B=Z X_1$. A
chain rule for the conditional entropy would mean that
\begin{align}
\h(X_0X_1|Z) = \h(X_1|Z) + \h(X_0|Z,X_1)\ .
\end{align}
Using Eq.~\eqref{eq:removeX1}, together with
Eqs.~\eqref{eq:subaddRHS} and~\eqref{eq:subaddLHS} again gives us a
contradiction. Note that $\pht(\cdot|\cdot)$ obeys conditions \{1\}
and \{2\}, and hence does not admit a chain rule in box world.

As $\ph(\cdot|\cdot)$ satisfies a chain rule, it follows from the
above that it must be `unreasonable'. Indeed, this can be seen from
the fact that $\ph(X_0|X_1Z)=1$ despite the fact that we can
perfectly determine the output of $X_0$ given $Z$ and $X_1$,
violating condition \{1\}.
It is easy to see that if we were to drop the conditions that make an entropy `reasonable' but
simply assume that it is not subadditive, but we do enforce a chain rule, then conditioning
can increase entropy.

\section{Information Causality}\label{sec:infoCausality}

We now use our entropic quantities to investigate the game given
in~\cite{infoCausality}. This task relates to `information
causality', which is expressed as the principle that `communication
of $k$ classical bits causes information gain of at most $k$ bits'.
In~\cite{infoCausality} it is reported that this principle can be
violated in box world using the following simple game (where we take
$k=1$): Alice is given two random classical bits $a_0$ and $a_1$ and
Bob is given a single random bit $t$. Alice is allowed to send a
single bit message $m$ to Bob, after which he must output a bit $b$.
The couple succeed in the task if $b=a_t$.

This task is clearly very similar to the non-local game considered
in section \ref{sec:boxEntropy}. Indeed, any solution to the
previous problem can also be used to solve this one. Alice takes the
parity bit as $x=a_0 \oplus a_1$, then generates $x_0$ and
$x_1=x_0\oplus x$ as before. She sends the message $m=x_0 \oplus
a_0$ to Bob. Using the previous protocol, Bob generates $x_t$, and
then outputs $b=x_t \oplus m = a_t$.

In the context of this game, `information causality' is interpreted
as meaning that
\begin{align}\label{eq:infoBound}
\ii := \ii(a_0;b|t=0)+\ii(a_1;b|t=1) \leq 1.
\end{align}
where $\ii(\cdot;\cdot|\cdot)$ is the classical conditional mutual
information. This inequality is obeyed in quantum theory. However,
given the above argument it is clear that it can be violated in box
world, as Alice and Bob can achieve $\ii=2$.

Let us examine why~\eqref{eq:infoBound} fails in terms of our
general entropies. We consider the state just after Bob has received
the message from Alice, when she holds classical bits $A_0$ and
$A_1$, and Bob holds the classical message $M$ and his part of the
PR-box $Z$. This state is described by
\begin{equation} \label{eq:icstate}
P(a_0 a_1 m \zout|\zin) = \left\{ \begin{array}{ccl} \frac{1}{8} & :
& \zout = a_{\zin} \oplus m \\ 0 &:& \textrm{otherwise}
\end{array} \right.
\end{equation}
We can compute entropies explicitly in this case as in section
\ref{sec:boxEntropy}, and will obtain similar results. However,
\cite{infoCausality} also contains a proof of \eqref{eq:infoBound}
in quantum theory based on the quantum mutual information. It is
interesting to attempt to follow this proof using our general mutual
information $\ih$ (or $\iht$) to see where it fails.

The quantum proof relies on the chain rule for quantum mutual
information (which $\ih$ satisfies by definition)\footnote{This
chain rule can be expressed as $\ih(A;BC) = \ih(A;C) + \ih(A;B|C)$,
where $\ih(A;B|C) = \ph(A|C) - \ph(A|BC)$.}, positivity of the
mutual information (which is true for $\ih$ in box world due to the
subadditivity of $\ph$), and non-signalling (which is one of the
defining features of box world). However, the crucial step is a use
of the data processing inequality to deduce that
\begin{align}
\ih(A_0;A_1 M Z) \geq \ih(A_0;M Z)
\end{align}
Although it is very natural that `forgetting' $A_1$ can only
decrease the mutual information, this inequality is violated in box
world. Indeed, for the state \eqref{eq:icstate} we find
\begin{align}
\ih(A_0;A_1 M Z) = 0,\qquad \ih(A_0;M Z) = 1
\end{align}
This is again a consequence of the \emph{violation of strong
subadditivity} for $\ph$, which forms the key ingredient in
why~\eqref{eq:infoBound} can be violated in box world.

Although the violation of $\eqref{eq:infoBound}$ in box world, and
its validity in quantum theory, is a very interesting result, it is
interesting to consider whether this really implies that
communicating $k$ bits has caused an information gain of more than
$k$ bits. From the state \eqref{eq:icstate} it is easy to check that
\begin{equation} \label{eq:icinf}
\ih(A_0A_1;M Z) = \iht(A_0 A_1 ;M Z) =1 \leq 1
\end{equation}
hence under both these measures the total information about the
composite system $A_0 A_1$ has only increased by  one bit due to the
one bit classical message. We show in Section~\ref{sec:boxChain} in
the appendix that in box world we indeed have that given some
arbitrary system $Z$ held by Bob, the mutual information about a
classical string $A$ can never increase by more than the length of a
classical message $M$ that is transmitted. Furthermore, Bob can
extract only one of the two bits, either $A_0$ \emph{or} $A_1$, with
the help of the message as is indeed noted in~\cite{infoCausality}.
It is therefore arguable that the information gain of Bob is only
one bit. Perhaps `information causality' should be restated in a
clearer way, that more directly represents the form of
\eqref{eq:infoBound}. e.g. the principle that an $m$ bit classical
communication allows us to learn any one out of at most $m$ unknown
bits.

\section{A simple coding theorem}\label{sec:codingTheorem}
We now show that for some theories, the entropic quantity
$\ph(\cdot)$ has an appealing operational interpretation in
capturing our ability to compress information. Here, we will only
show this for theories obeying further restrictions, and it is an
interesting open question how generally this interpretation applies.

\subsection{Dimension and subspaces}
Before we can talk about compression, we first need to clarify our
notions of the size of a system.
Intuitively, the size of a system should limit the amount of uncertainty
we can have about it. Furthermore, to compress,
we will clearly need to shrink  the original state space.
It is therefore helpful to define a notion of size for
any subset of allowed states $\cS_T \subseteq \cS$.

We refer to the size of a set of states $\cS_T$ as its \emph{dimension} $d$,
which we define by
\begin{align}
 d:=\min_{\ef \in \cE^*}  | \{ r \in \cR_{\ef} | \exists S \in \cS_T, e_r(S)>0 \}|.
\end{align}
This corresponds to eliminating all measurement outcomes that cannot occur for any state in
$\cS_T$, and then counting the minimal number of remaining outcomes for any fine-grained measurement.
It follows that $\log d \geq \ph(S) $ for all $S \in \cS_T$.
In quantum theory $d$ corresponds
precisely to the dimension of a Hilbert space.

A natural way to select a subset of states is to consider all states
that yield a given measurement outcome with certainty. We refer to an effect $f$ such that
$\{f,u-f\}$ is an allowed measurement, and that occurs with certainty for some state, as a \emph{full} effect (i.e. $f$ is full if there exists $S\in \cS$ such that $f(S)=1$).
For any full effect $f$, we can therefore define a non-empty
subset of states $\cS_f = \{ S| S \in \cS, f(S)=1\}$. We refer to
such a subset as the \emph{subspace} of $\cS$ given by $f$. Note
that subspaces are always convex, and the subspace corresponding to an effect $f$ which is both full \emph{and}
fine-grained obeys $d_f=1$.

We say that we have compressed a state if we have constrained it to lie
within a set of states of smaller dimension.

\subsection{Additional assumptions}
So far, we were never concerned about what happens to a state after
a measurement. In our compression protocol, however, we will need to
use an abstract notion of post-measurement states as described in
Section~\ref{sec:trans}.  In particular, we will consider
\emph{pseudo-projective measurements}, which we define to be
measurements that fullfill two conditions.
\begin{enumerate}
\item \emph{(Repeatability)} \label{ass:repeatability}
A pseudo-projective measurement is repeatable, such that if the same
measurement is applied again the same result is obtained. This
requires that the output state $S_r$ after obtaining a result $r$
lies in the subspace given by $e_r$ (i.e., $e_r(S_r)=1$).
Consequently, all effects in a pseudo-projective measurement must be
full effects.
\item \emph{(Weak Disturbance)}
If a particular outcome $r$ of a pseudo-projective measurement
occurs with probability $e_r(S) \geq 1-\delta$ for a state $S$, then
the post measurement state $S_r$ after this result is obtained
satisfies $e_r(S) \dist(S,S_r) \leq c \delta^\eps$, where $c \geq 0$
and $\eps \in (0,1]$ are constants depending on the particular
theory. For example, for projective measurements in quantum theory
$c = (\sqrt{8}+1)/2$ and $\eps = 1/2$.
\end{enumerate} Any projective
measurement in quantum theory fulfills these conditions, but these
conditions alone do not define projective measurements, hence the
slightly different name. In quantum theory, the weak disturbance
property can be understood as an instance of the gentle measurement
lemma~\cite{andreas:gml}.

Furthermore, in order to prove our simple coding theorem, we will
need to make some additional assumptions on the states and the
measurements that achieve the minimal output entropy $\ph(\cdot)$ in
our theory. In particular, we assume that for all states, the
minimal output entropy can be attained by a pseudo-projective
measurement. That is, we assume that for all $S \in \cS$ there
exists some pseudo-projective measurement $\mbf{e} \in \cE^*$ such
that $\ph(S) = H(\mbf{e}(S))$. We further assume that for all such
measurements, $\mbf{e}^{\otimes n}$ is fine-grained and
pseudo-projective, and that course grainings of $\mbf{e}^{\otimes
n}$ can also be made pseudo-projective. Lastly, we assume that the
dimension of $\cS^{\otimes n}$ is $d^n$. These assumptions are all
true in the classical and quantum case (where $\mbf{e}$ is
projective).

We will see in Appendix~\ref{sec:codingTheoremProof},
that this is all we will need to show the following simple coding
theorem following the steps taken by Shannon~\cite{shannon:entropy}
and Schumacher~\cite{schumacher:coding} (see for
example~\cite{nielsen&chuang:qc}).

\subsection{Compression}

We consider a source that emits a state $\tilde{S}_k \in \cS$
with probability $q_k$, chosen independently at random in each time
step. When considering $n$ time steps, we hence obtain a sequence of
states $\tilde{S}_{\vec{k}} = \tilde{S}_{k_1},\ldots,\tilde{S}_{k_n}
\in \cS^{\otimes n}$ with $\vec{k} = (k_1,\ldots,k_n)$, where each
sequence occurs with probability $q_{\vec{k}} = \Pi_j q_{k_j}$. A
compression scheme consists of an encoding and decoding procedure.
The encoding procedure maps each possible $\tilde{S}_{\vec{k}}$ into
a state $\hat{S}_{\vec{k}} \in \cS_f \subset \cS^{\otimes n}$.
In turn the decoding procedure maps the states $\hat{S}_{\vec{k}}$ back to states
$\breve{S}_{\vec{k}} \in \cS$ on the original state space. In
analogy with the quantum case, we say that the \emph{compression
scheme has rate} $R$, if the dimension of the smaller space obeys
$d_f \leq 2^{nR}$. Note that in order for a compression scheme to be
useful, it must have $R<\log d$  (and hence $d_f <d^n$).
A compression scheme is called \emph{reliable}, if we can
recover the original state (almost)
perfectly, in the sense that the average distance between the
original and the reconstructed state can be made arbitrarily small
for sufficiently large $n$. I.e. for any $\epsilon>0$ and all sufficiently large $n$,
\begin{align}
\sum_k q_{\vec{k}} \dist(\tilde{S}_{\vec{k}},\breve{S}_{\vec{k}})
\leq \epsilon\ .
\end{align}

Note that the output of the source can be described as a mixed state
$\src = \sum_k q_k \tilde{S}_k$ in each time step, and a product
state $\src^{\otimes n} \in \cS^{\otimes n}$ over the course of $n$
time steps. We then obtain the following theorem (see appendix Section~\ref{sec:codingApp}) in terms of the
entropy of the source $\ph(\src)$.

\begin{theorem}\label{thm:codingTheorem}
Consider an i.i.d source $\{q_k,\tilde{S}_k \in \cS\}_k$ with
entropy rate $\ph(\src)$. Then for $R > \ph(\src)$ there exists a
reliable compression scheme with rate $R$.
\end{theorem}

Note that in order to establish that $\ph(\cdot)$ truly characterizes our ability to compress information,
we would also like to have a converse stating that for $R < \ph(\src)$ there exists no reliable compression scheme.
In quantum theory, it is not hard to prove the converse of the above theorem since it admits a strong duality
between states and measurements, which may also hold for other theories. Here, however, we explicitly tried
to avoid introducing any such strong assumptions.

\section{Conclusion and open questions}\label{sec:openQuestions}

We introduced entropic measures to quantify information in any
physical theory that admits minimal notions of systems, states and
measurements. Even though these measures necessarily have some
limitations, we nevertheless showed that they also exhibit many
intuitive properties, and for some theories have an appealing
operational interpretation, quantifying our ability to compress
states. Most of the problems we encountered with the conditional
entropy seem to arise due to a violation of strong subadditivity. It
is an interesting question whether quantum and classical theories
are the only ones in which $\ph$ is strongly subadditive, or whether
this is true for other theories. Indeed, it would be an exciting
question to turn things around and start by demanding that our
entropic measures \emph{do} satisfies these properties, and
determine how this restricts the set of possible theories.

In $\pht(\cdot|\cdot)$ we defined a natural entropic quantity which
differs from the conditional von Neumann entropy in quantum theory, and has
been used in~\cite{discord} to study quantum correlations.
It would be interesting to study whether this quantity can shed any further
light on quantum phenomena, or if an alternative conditional entropy
can be defined that behaves like $\pht(\cdot|\cdot)$ in box world,
but still reduces to the conditional shannon entropy in quantum
theory.

Whereas we have proved some intuitive properties of our quantities,
it is interesting to see whether other properties of the von Neumann
or Shannon entropy carry over to this setting. In particular, it
would be interesting to prove bounds on the mutual and accessible
information analogous to Holevo's theorem when none of the systems
are classical.

Another interesting question is whether one can find a closed form
expression for the relative entropy in general theories. In quantum
theory, we can define the mutual information (and indeed the entropy
itself) in terms of the relative entropy \footnote{In particular,
the mutual information for a quantum state $\rho_{AB}$ is the same
as the relative entropy between $\rho_{AB}$ and $\rho_A \otimes
\rho_B$, and the entropy of $\rho$ is (minus) the relative entropy
between $\rho$ and the identity operator.}, hence such an approach
may also yield an alternative definition of other entropic
quantities for general theories.

We believe our measures are an interesting step towards
understanding information processing in general physical theories,
which may in turn shed some light on our own quantum world.

\acknowledgments

The non-local game used in our example above was discovered in
collaboration with Andrew Doherty, whom we thank for the kind
permission to use it here. The authors also thank Sergio Boixo,
Matthew Elliot and Jonathan Oppenheim for interesting discussions,
and Matt Leifer and Ronald de Wolf for comments on an earlier draft.
SW is supported by NSF grants PHY-04056720 and PHY-0803371. AJS is
supported by a Royal Society URF, and in part by the EU QAP project
(CT-015848). Part of this work was done while AJS was visiting
Caltech (Pasadena, USA).

\emph{Note added:} In the course of this work we learned of independent
work on the same general topic~\cite{hb:entropy}, to appear
simultaneously in NJP. Related work has also appeared later on~\cite{gk:entropy}.

\appendix

\setcounter{section}{0}
\smallskip
In this appendix, we provide formal
statements and the technical details of our claims.

\section{Distance metric} \label{app:distance}

We now show that the quantity~\eqref{eq:distanceMeasure} is indeed a metric on the state space $\cS$.
\begin{lemma}\label{lem:distanceMeasure}
$\dist: \cS \times \cS \rightarrow [0,1]$ as defined in~\eqref{eq:distanceMeasure} is a metric on the state space $\cS$.
\end{lemma}
\begin{proof}
Consider states $S_0,S_1,S_2 \in \cS$.
Clearly,
\begin{align}
\dist(S_0,S_1) \geq 0
\end{align}
using the property of the classical statistical distance,
where equality holds iff $S_0 = S_1$ by definition of the state space $\cS$. It remains to show
that $\dist$ obeys a triangle inequality. Let $\ef_{ij}$ be the optimal measurement to distinguish states $i$ and $j$.
We then have
\begin{align}
\dist(S_0,S_1)& + \dist(S_1,S_2)\\
&\geq \cdist(\ef_{02}(S_0),\ef_{02}(S_1)) + \cdist(\ef_{02}(S_1),\ef_{02}(S_2)) \nonumber\\
&\geq \cdist(\ef_{02}(S_0),\ef_{02}(S_2)) = \dist(S_0,S_2)\ , \nonumber
\end{align}
where the second inequality follows from the fact that the classical statistical distance $C$ itself obeys
the triangle inequality.
\end{proof}

\section{Properties of $\ph$} \label{app:entropy}

In this appendix we derive properties of the entropy $\ph$ used in
the paper. Note that by assumption $\mathcal{E}^*$ is non-empty,
which implies that $\ph(S)$ is well-defined.

\subsection{Reduction to the von Neumann and Shannon entropy}
\label{app:entropyReduction}

We now show that the entropic quantity~\eqref{eq:entropyDef1}
reduces to the von Neumann and Shannon entropy in the classical and
quantum settings respectively. For the relation to the von Neumann
entropy, we will need the following little lemma.
\begin{lemma}\label{lem:toVonNeumannStep1}
Let $\rho \in \bop(\hil)$ be a quantum state with eigendecomposition
$\rho = \sum_j p_j \proj{\psi_j}$. Then
\begin{align}
\ph(\rho) = \nh(\rho) = \sh(\vec{p})\ ,
\end{align}
where $\vec{p} = (p_1,\ldots,p_d)$ with $d = \dim(\hil)$.
\end{lemma}
\begin{proof}
Our goal will be to show that for any fine-grained measurement
$\mbf{e}$ with
\begin{align}\label{eq:extremalOther}
e_l &=  c_\ell \proj{\phi_\ell} \in \bop(\hil)\ \mid\  0 \leq c_\ell \leq 1\\
&\qquad \mbox{ and } \sum_{\ell} c_\ell \proj{\phi_\ell} = \id
\nonumber
\end{align}
the Shannon entropy of the distribution $ q_\ell := c_\ell
\bra{\phi_\ell}\rho \ket{\phi_\ell} $ is always at least as large as
the distribution obtained by measuring in the eigenbasis of $\rho$,
that is,
\begin{align}
\sh(\vec{p}) \leq \sh(\vec{q})\ ,
\end{align}
with $\vec{q} = (q_1,\ldots,q_N)$.

Let $N = |\ef|$ and note that $d \leq N$. First of all, note that we
can always extend a distribution $\{p_j\}$ over $d$ elements to a
distribution $\{\tilde{p}_j\}$ over $N$ elements by letting
$\tilde{p}_j = p_j$ for all $j \leq d$ and $\tilde{p}_j = 0$ for all
$j
> d$. Clearly, $\sh(\vec{\tilde{p}}) = \sh(\vec{p})$ with
$\vec{\tilde{p}} = (\tilde{p}_1,\ldots,\tilde{p}_N)$.

Second, note that
\begin{align}
q_\ell = \sum_j p_j q_{\ell|j} &\qquad\mbox{ and }\qquad q_{\ell|j}
= c_\ell |\inp{\phi_\ell}{\psi_j}|^2\ ,
\end{align}
from which we immediately obtain together
with~\eqref{eq:extremalOther} that
\begin{align}
\sum_{j} q_{\ell|j} = c_\ell&\qquad \mbox{ and }\qquad \sum_{\ell}
q_{\ell|j} = 1\ .
\end{align}
Consider the $N\times N$ matrix $M$ determined by the entries
\begin{align}
M_{\ell,j} = \left\{\begin{array}{ll}
q_{\ell|j}& \mbox{ for } j \leq d\, \\
\frac{1-c_\ell}{N-d}& \mbox{ for } j > d\ .
\end{array}\right.
\end{align}
which allows us to write $\vec{q} = M \vec{\tilde{p}}$. Note that
since $M_{\ell,j} \geq 0$ and $\sum_j M_{\ell,j} = \sum_\ell
M_{\ell,j} = 1$, $M$ is a doubly stochastic matrix. Using Birkhoff's
theorem (see e.g .,~\cite[Theorem 8.7.1]{horn&johnson:ma}), we may
thus write $M$ as a convex combination of permutation matrices, that
is,
\begin{align}
M = \sum_{\pi \in S_N} P(\pi) \pi\ ,
\end{align}
where $P$ is a probability distribution over the group of
permutations $S_N$. Using the concavity of the Shannon entropy we
obtain
\begin{align}
\sh(\vec{q}) \geq \sum_{\pi \in S_N} P(\pi)
\sh(\pi(\vec{\tilde{p}})) = \sh(\vec{p})\ .
\end{align}
As we can always measure $\rho$ in its eigenbasis it follows that
\begin{align}
\ph(\rho) =\inf_{\mbf{e} \in \cE^*} \sh(\mbf{e}(S)) = \inf_{\vec{q}}
\sh(\vec{q}) = \sh(\vec{p}).
\end{align}
and it is easy to see that $\sh(\vec{p}) = \nh(\rho)$.
\end{proof}

Since the von Neumann entropy reduces to the Shannon entropy in a
classical setting, this also shows that the entropic
quantity~\eqref{eq:entropyDef1} reduces to the Shannon entropy in
the classical case.

\subsection{Positivity, Boundedness, and Concavity} \label{app:entropyOther}

Here we prove the other general properties of the entropy $\ph$.

\emph{Positivity:} This follows trivially from the Positivity of the Shannon
Entropy.

\emph{Boundedness:} The existence of a measurement $\ef\in \mathcal{E}^*$ with $d$
outcomes, combined with the fact that the Shannon entropy is
maximized for a uniform probability distribution, ensure that
\begin{align}
\ph(S) \leq \sh(\ef(S)) \leq \log(d)
\end{align}
which gives Boundedness.

\emph{Concavity:} To see that $\ph$ is concave, suppose first that the infimum in the
definition~\eqref{eq:entropyDef1} of $\ph(S_{\textrm{mix}})$ is
achieved, such that $\ph(S_{\textrm{mix}}) =
\sh(\ef(S_{\textrm{mix}}))$ for some $\ef \in \cE^*$. As effects are
linear maps, $\ef(S_{\textrm{mix}}) = p \ef(S_1) + (1-p) \ef(S_2)$.
Hence, by the concavity of the Shannon entropy
\begin{eqnarray}
\ph(S_{\textrm{mix}}) &=& \sh(\ef(S_{\textrm{mix}})) \\
&\geq& p \sh(\ef(S_1)) + (1-p) \sh(\ef(S_2))\nonumber \\
&\geq& p \ph(S_1) + (1-p) \ph(S_2) \nonumber
\end{eqnarray}
which concludes our claim. On the other hand, if the infimum is not
achievable then for all sufficiently small $\delta>0$ we can find an
$\ef \in \cE^*$ such that $\ph(S_{\textrm{mix}}) =
\sh(\ef(S_{\textrm{mix}})) - \delta$. Using the same argument as
before, we find
\begin{align}
\ph(S_{\textrm{mix}}) \geq p \ph(S_1) + (1-p) \ph(S_2) - \delta
\end{align}
As this holds for all sufficiently small $\delta$ the result
follows.

\subsection{Limited Subadditivity and Continuity}

Here we prove two properties of $\ph$ that require additional minor
assumptions on our theory. However, they are obeyed in quantum
theory, classical theory and box world.

\emph{Limited Subadditivity:} Given an additional reasonable
assumption, we can prove that $\ph$ is subadditive, we first assume
that there exist $\ef \in \mathcal{E}_A^*$ and $\fef \in
\mathcal{E}_B^*$ such that $\ph(A) = \sh(\ef(A))$ and $\ph(B) =
\sh(\fef(B))$. By assumption, $\ef \otimes \fef$ is a fine-grained
measurement on the joint system $\mathcal{AB}$. Thus by the
subadditivity of the Shannon entropy
\begin{eqnarray}
\ph(A) + \ph(B) &=& \sh(\mbf{e}(A)) + \sh(\mbf{f}(B))
\\
&\geq& \sh((\ef \otimes \fef)AB) \nonumber\\
&\geq& \ph(AB)\ ,\nonumber
\end{eqnarray}
as claimed. Now suppose that the infimum for one or both of $\ph(A)$
or $\ph(B)$ is not achieved. Then for all sufficiently small
$\delta>0$ we can find $\ef \in \mathcal{E}_A^*$ and $\fef \in
\mathcal{E}_B^*$ such that
\begin{align}
\ph(A) + \ph(B) = \sh(\mbf{e}(A)) + \sh(\mbf{f}(B)) - \delta \geq
\ph(AB) - \delta.
\end{align}
As this holds for all sufficiently small $\delta>0$ the result
follows.

Note that if $A$ and $B$ are in a product state, and the theory only
allows product measurements on $AB$ then equality holds
in~\eqref{eq:subadd}. However given we allow an arbitrary set of
joint measurements, equality does not hold when $A$ and $B$ are in a
product state for any possible probabilistic theories (Consider the
case in which $\ph(A)>\log 2$, but there exists a fine-grained
measurement on $AB$ with only 2 outcomes).

\emph{Limited Continuity:} Here we prove an analogue of the Fannes
inequality~\cite{fannes:inequ}, given an additional reasonable
assumption that we can restrict to measurements with at most $D$
outcomes without changing the entropy of a system.

Suppose without loss of generality that $\ph(S_1) \geq \ph(S_2)$.
Initially, we also suppose that the infimum in the definition of
$\ph(S_2)$ is achieved for some $\fef \in \cE^*$, such that $\ph(S_2) =
\sh(\fef(S_2))$. We can then bound
\begin{align}
|\ph(S_1) - \ph(S_2)| &\leq |\sh(\fef(S_1)) - \sh(\fef(S_2))|\\
&\leq \cdist(\fef(S_1),\fef(S_2)) \log \left( \frac{D}{\cdist(\fef(S_1),\fef(S_2))}\right) \nonumber\\
&\leq \dist(S_1,S_2) \log \left( \frac{D}{\dist(S_1,S_2)}\right) \nonumber
\end{align}
where the first inequality follows from the fact that $\ph(S_1) \leq \sh(\fef(S_1))$, the second
from Fannes inequality~\cite{fannes:inequ} applied to the classical case, and the final inequality
by noting that
\begin{align}
\cdist(\fef(S_1),\fef(S_2)) \leq \dist(S_1,S_2)\ < \frac{1}{e},
\end{align}
If the infimum is not achieved, then for all sufficiently small
$\delta>0$ there nevertheless exists $\fef \in \cE^*$ such that
$\ph(S_2) = \sh(\fef(S_2)) - \delta$. Following the same procedure as
before, we find
\begin{align}
|\ph(S_1) - \ph(S_2)| \leq \dist(S_1,S_2) \log \left(
\frac{D}{\dist(S_1,S_2)}\right) + \delta
\end{align}
from which the result follows.

\section{Properties of the conditional entropy} \label{app:condEntropy}

\subsection{General case}
We now show that in contrast to the quantity $\ph$, our second form
of conditional entropy $\pht$ obeys the intuitive property that
conditioning reduces entropy in all cases.

\begin{lemma}[Conditioning reduces entropy for $\pht$]\label{lem:conditioningReducesEntropy}
For any tripartite state $ABC \in \cS_{ABC}$ and its corresponding
reduced states we have
\begin{align}
\pht(A) \geq \pht(A|B) \geq \pht(A|BC) \ .
\end{align}
\end{lemma}
\begin{proof}
The first inequality follows by choosing the unit measurement in the
infimum over $\mathcal{E}_B$ in the definition of $\pht(A|B)$, and
noting that $\ph(A) = u(B) \ph(A_{|u}) \geq \ph(A|B)$. The second
inequality comes from restricting to measurements of the form
$\mbf{f}_B \otimes \mbf{u}_C$ in the infimum over $\mathcal{E}_{BC}$
in the definition of $\pht(A|BC)$.
\end{proof}

\subsection{Box world}\label{sec:boxChain}

We now prove a very restricted form of chain rule in box world. This will allow us to show that for our notions
of entropy the mutual information about any classical information given an arbitrary state in box world can never
increase by more than $\ell$ bits when transmitting $\ell$ bits of information. To show our simple chain rule, we will use
the fact that in box world, we have that when considering a composite of a classical system $M$ and an arbitrary system $B$,
the only allowed measurements on the composite system $MB$ take the form of first performing the only allowed measurement on $M$,
followed by a choice of measurement on $B$ that may depend on the outcome of the measurement on $M$.
Since classical systems in box world admit exactly one measurement (possibly followed by
some classical post-processing), we simply write $\sh(M) = \ph(M)$ to denote
the resulting entropy.

\begin{lemma}[Box chain rule]\label{lem:conditionalChain}
For any tripartite state $CMB \in \cS_{CMB}$ in box world, where its corresponding reduced states
where $C$ and $M$ are classical we have
\begin{align}
\pht(C|MB) \geq \pht(CM|B) - \ph(M)\ .
\end{align}
\end{lemma}
\begin{proof}
For simplicity, we only examine the case where the infimum is attained in $\pht$, the other case can again be obtained
by taking the appropriate limit.
Since the only measurements on $MB$ are as described above, we clearly have
\begin{align}
\pht(C|MB) &= \sum_{m} e_{m}(M) \sum_k f_k(B_{|m}) \pht(C_{|m,k}) =\\
&=\sum_{m} e_{m}(M) \sum_k f_k(B_{|m}) \sh(C|M = m, K = k)\nonumber\\
&=\sh(C|M,K) = \sh(C M|K) - \sh(M|K)\nonumber\\
&\geq \pht(C M|B) - \ph(M)\nonumber\ ,
\end{align}
where the first equality follows from the definition of $\pht$ and
the fact that $M$ is classical, the second from the definition of
the conditional Shannon entropy, the third from the chain rule for
the conditional Shannon entropy, and the final inequality from the
definition of $\pht$, the fact that $\ph(M) = \sh(\ef(M))$ for
classical systems and the fact that conditioning reduces entropy for
the Shannon entropy.
\end{proof}

We now see that in consistency with the no-signalling principle, the transmition of an $\ell$
bit message $M$ causes the mutual information about a classical system $C$ given access to some aribtrary
box information $B$ to increase by at most $\ell$ bits.
Note that for our alternate definition of conditional entropy and mutual information
we have
\begin{align}
\iht(C;M B) = \ph(C) - \pht(C|MB)\ .
\end{align}
First, note that we can write
\begin{align}
\iht(C;MB) &= \iht(C;B) + \iht(C;M|B)\ ,\\
\iht(C;B) &= \ph(C)- \pht(C|B)\ ,\nonumber\\
\iht(C;M|B) &:= \pht(C|B) - \pht(C|MB)\ , \nonumber
\end{align}
by definition.
We hence have
\begin{align}
\iht(C;MB) &\leq \pht(C|B) + \pht(M) - \pht(CM|B)\\
&\leq \iht(C;B) + \pht(M) \leq \iht(C;B)+ \ell\ .\nonumber
\end{align}

\section{Properties of $\bh$} \label{app:decomposition}

In this section we explore properties of the decomposition entropy
$\bh$.

\subsection{Reduction to the von Neumann and Shannon entropy}

To show the reduction of $\bh(\rho)$ to the von Neumann
entropy $S(\rho)$ in quantum theory, we use the following Lemma

\begin{lemma}[Theorem 11.10 in~\cite{nielsen&chuang:qc}] Suppose $\rho =
\sum_i p_i \rho_i$, where $p_i$ are some set of probabilities and
$\rho_i$ are density operators. Then
\begin{align}
\sv(\rho) \leq \sum_i p_i \sv(\rho_i) + \sh(p_i),
\end{align}
with equality if and only if the states $\rho_i$ have support on
orthogonal subspaces.
\end{lemma}

Note that when $\rho_i$ are pure states, $\sv(\rho_i)=0$. Hence for
any pure state decomposition $\mbf{D}(\rho)$, this implies
\begin{align}
\sv(\rho) \leq \sh(\mbf{D}(\rho))
\end{align}
Furthermore, denoting an eigendecomposition of $\rho$ by
$\mbf{D}^{*}(\rho)$, it is easy to see that $\sh(\mbf{D}^*(\rho)) =
\sv(\rho)$. Hence it follows that
\begin{align}
\sv(\rho) = \bh(\rho) = \inf_{\mbf{D}(\rho)} \sh(\mbf{D}(\rho))
\end{align}

\subsection{Subadditivity and concavity}

In this section we will show that $\bh$ is neither concave nor
subadditive by giving explicit counterexamples from box world.

First consider a single box with binary input/output. For clarity,
we will represent its state by giving its probability distribution
$P(a|x)$ in vector form:
\begin{align}
S = \left( \begin{array}{c} P(0|0) \\ P(1|0) \\ \hline P(0|1) \\
P(1|1) \end{array} \right)
\end{align}
Now consider the two states
\begin{align}
S_1 =\left( \begin{array}{c} 1 \\ 0 \\ \hline 1/2 \\
1/2 \end{array} \right), \qquad
S_2 =\left( \begin{array}{c} 1/2 \\ 1/2 \\ \hline 1 \\
0 \end{array} \right),
\end{align}
These can both be optimally decomposed into two equally weighted
pure states, e.g.
\begin{align}
S_1 = \frac{1}{2}\left( \begin{array}{c} 1 \\ 0 \\ \hline 1 \\
0 \end{array} \right) + \frac{1}{2} \left( \begin{array}{c} 1 \\ 0 \\ \hline 0 \\
1 \end{array} \right)
\end{align}
hence they satisfy $\bh(S_1) = \bh(S_2) = \log 2 = 1$.
However now consider the mixed state,
\begin{align}
S_{\textrm{mix}} = \frac{1}{2} S_1 + \frac{1}{2} S_2 =\left( \begin{array}{c} 3/4 \\ 1/4 \\ \hline 3/4 \\
1/4 \end{array} \right) = \frac{1}{4}\left( \begin{array}{c} 1 \\ 0 \\ \hline 1 \\
0 \end{array} \right) + \frac{3}{4}\left( \begin{array}{c} 0 \\ 1 \\ \hline 0 \\
1 \end{array} \right).
\end{align}
which has $\bh(S_{\textrm{mix}})=\sh((\frac{3}{4}, \frac{1}{4}))
<1$. Hence in this case we violate concavity
\begin{align}
\bh(S_{\textrm{mix}}) <\frac{1}{2} \bh(S_1) + \frac{1}{2}
\bh(S_2).
\end{align}

To obtain a violation of subadditivity we consider a bipartite state
in which each system has a binary input/output, represented in the
form of a matrix
\begin{align}
S_{AB} = \left( \begin{array}{cc|cc} P(00|00) & P(01|00) & P(00|01)
& P(01|01) \\ P(10|00) & P(11|00) & P(10|01) & P(11|01) \\ \hline
P(00|10) & P(01|10) & P(00|11) & P(01|11) \\ P(10|10) & P(11|10) &
P(10|11) & P(11|11)
\end{array} \right)
\end{align}
Choose the following allowed state
\begin{align}
S_{AB} = \frac{1}{8} \left( \begin{array}{cc|cc} 2 & 3 &
2 & 3 \\ 3 & 0 & 3 & 0 \\
\hline 5 & 0 & 2 & 3 \\ 0 & 3 & 3 & 0
\end{array} \right), \qquad A=B= \frac{1}{8}\left( \begin{array}{c} 5 \\ 3 \\ \hline 5 \\
3 \end{array} \right).
\end{align}
It is known that in this case there are exactly 24 pure states for
the bipartite binary input/output case (16 product states and 8
entangled states)~\cite{barrett}, which we denote by $S_{AB}^i$. By
demanding that $S_{AB} - p_i S_{AB}^i$ be a positive matrix for each
pure state we find that any decomposition must satisfy $p_i \leq
\frac{1}{4} \, \forall \,i$. Hence $\bh(AB)=\inf_{\mbf{D}(\rho)}
\sh(p_i) \geq 2$. In fact we can construct an explicit decomposition
in terms of an entangled state and three product states (all equally
weighted), giving $\bh(AB)=2$. The marginal states on the other hand
satisfy
\begin{align}
\bh(A) = \bh(B) = \sh\left( \left( \frac{3}{8},
\frac{5}{8} \right)\right) < 1.
\end{align}
Hence we obtain
\begin{align}
\bh(AB) >\bh(A) + \bh(B)
\end{align}
in violation of subadditivity.

\section{A simple coding theorem}\label{sec:codingTheoremProof}\label{sec:codingApp}

We now sketch the proof of Theorem~\ref{thm:codingTheorem}, which is
straightforward following the steps taken in the quantum
setting~\cite{schumacher:coding}.

Consider the pseudo-projective measurement $\ef$ that gives the
minimal output entropy for the state $\src$, which we take to exist
by assumption.

At the core of our little coding theorem lies an observation about
$\eps$-typical sequences analogous to the classical and quantum
setting. Define the set of $\eps$-typical outcomes when measuring
$\ef^{\otimes n}$ on the state $\src^{\otimes n} \in \cS^{\otimes
n}$ as
\begin{align}
T(n,\eps) &:= \left\{r_1,\ldots,r_n \in \cR_{\ef}^{\times n}\right.\\
 &\left.\mid \left|\frac{1}{n} \log\left(\frac{1}{e_{r_1}(\src)\ldots e_{r_n}(\src)}\right) - \ph(\src)\right| \leq \eps\right\}\ .\nonumber
\end{align}
When $n$ and $\eps$ are clear from context, we will also use the effects
\begin{align}
h_T(\src^{\otimes n}) &:= \sum_{\vec{r} \in T(n,\eps)} e_{\vec{r}}(\src^{\otimes n})\ ,\\
h_A &:= u - h_T\ .
\end{align}
Since we assumed that any theory contains arbitrary coarse-grainings of measurements, we can consider the measurement
\begin{align}\label{eq:projTypical}
\hef &:= \{(T,h_T),(A,h_A)\}\ ,
\end{align}
which by assumption we can make pseudo-projective. We refer to the
subspaces given by $h_T$ and $h_A$ as the typical and atypical
subspaces respectively. If we observe outcome 'T' for the
measurement $\hef$, we conclude that a state lies in the typical
subspace associated with the set $T(n,\eps)$. Otherwise, we conclude
that the states lies in the atypical subspace.

Note that by assumption we have that $\ef^{\otimes n}$ is a
fine-grained measurement. For all states in the typical subspace,
only outcomes in the typical set $T(n,\eps)$ will occur. Hence we
have that the dimension of the typical subspace satisfies $d_T  \leq
|T(n,\eps)| $.

We are now ready to prove the following theorem:
\begin{theorem}[Typical subspace theorem]
Let all quantities be defined as above.
Fix $\eps > 0$, then for any $\delta > 0$ and sufficiently large $n$,
\begin{align}
&(i)\qquad h_T(\src^{\otimes n}) \geq 1 - \delta\ .
\end{align}
\begin{align}
&(ii)\qquad (1-\delta) 2^{n(\ph(\src)-\eps)} \leq |T(n,\eps)| \leq
2^{n(\ph(\src) + \eps)}\ .
\end{align}
\end{theorem}
\begin{proof}
The proof of (i) and (ii) is analogous to~\cite[Theorem 12.5]{nielsen&chuang:qc} by noting that
\begin{align}
h_T(\src^{\otimes n}) &= \sum_{(r_1,\ldots,r_n) \in T(n,\eps)} e_{r_1}(\src)e_{r_2}(\src)\ldots e_{r_n}(\src)\ ,
\end{align}
and that the condition characterizing the set $T(n,\eps)$ of $\eps$-typical sequences can also be written
as
\begin{align}
2^{-n(\ph(\src) + \eps)} \leq e_{r_1}(\src)\ldots e_{r_n}(\src) \leq
2^{-n (\ph(\src) - \eps)}\ .
\end{align}

Given the statement about typical sequences, we can now complete the
proof of Theorem~\ref{thm:codingTheorem}: Recall that the source
emits a sequence of states $\tilde{S}_{\vec{k}}$ with probability
$q_{\vec{k}}$. To compress the state we perform a pseudo-projective
measurement of $\hef$ given by~\eqref{eq:projTypical}. If we obtain
outcome `T' (corresponding to the typical subspace) we output the
post-measurement state $T[\tilde{S}_{\vec{k}}]$, which must lie in
the typical subspace as the measurement is repeatable. Otherwise, we
prepare an arbitrary fixed state in the typical subspace which we
will call $S_{\rm fail}$. The resulting state is thus a mixed state
in the typical subspace of the form
\begin{align}
\hat{S}_{\vec{k}} = h_T(\tilde{S}_{\vec{k}}) T[\tilde{S}_{\vec{k}}]
+ h_A(\tilde{S}_{\vec{k}}) S_{\rm fail}\ .
\end{align}
Note that condition (ii) of the theorem tells us that the dimension
of the typical subspace is at most $2^{n(\ph(\src) + \eps)}$. For
any $R>\ph(\src)$, we can therefore find an $\eps$ such that we
achieve a compression of rate $R$.

To decompress, we will do nothing and simply output
\begin{align}
\breve{S}_{\vec{k}} := \hat{S}_{\vec{k}}\ ,
\end{align}
and so all that remains is to show that $\hat{S}_{\vec{k}}$ is in
fact close to the original state $\tilde{S}_{\vec{k}}$. Suppose for
simplicity that the maximum is attained when computing the distance,
and let $\ef$ denote the optimal measurement. That is
\begin{align}
\dist(\hat{S}_{\vec{k}}, \tilde{S}_{\vec{k}}) &= \sup_{\fef}
\cdist(\fef(\hat{S}_{\vec{k}}),\fef(\tilde{S}_{\vec{k}})) =
\cdist(\ef(\hat{S}_{\vec{k}}),\ef(\tilde{S}_{\vec{k}})),
\end{align}
 We then have
\begin{align}
\dist(\breve{S}_{\vec{k}}, \tilde{S}_{\vec{k}}) &=  \cdist(\ef(\hat{S}_{\vec{k}}),\ef(\tilde{S}_{\vec{k}}))\\
&\leq  h_T(\tilde{S}_{\vec{k}})
\cdist(\ef(T[\tilde{S}_{\vec{k}}]),\ef(\tilde{S}_{\vec{k}})) + \nonumber \\
 &\qquad h_A(\tilde{S}_{\vec{k}})
\cdist(\ef(S_{\rm fail}),\ef(\tilde{S}_{\vec{k}})) \nonumber \\
&\leq  h_T(\tilde{S}_{\vec{k}})
\dist(T[\tilde{S}_{\vec{k}}],\tilde{S}_{\vec{k}}) + \nonumber \\
 &\qquad h_A(\tilde{S}_{\vec{k}})
\dist(S_{\rm fail},\tilde{S}_{\vec{k}})  \nonumber \\
& \leq c h_A(\tilde{S}_{\vec{k}})^{\eps} + h_A(\tilde{S}_{\vec{k}})\ , \nonumber \\
& \leq (c+1) h_A(\tilde{S}_{\vec{k}})^{\eps} \nonumber
\end{align}
where the first inequality follows from the properties of the
classical trace distance and the linearity of effects, the second
from the definition of distance, and the third from the weak
disturbance property of a pseudo-projective measurement, where $c
\geq 0$ and $\eps \in (0,1]$ are constants given by a particular
theory. We then note that
\begin{align}
\sum_{\vec{k}} q_{\vec{k}}
\dist(\tilde{S}_{\vec{k}},\breve{S}_{\vec{k}}) &\leq \left(
\sum_{\vec{k}} q_{\vec{k}}
\dist(\tilde{S}_{\vec{k}},\breve{S}_{\vec{k}})^{\frac{1}{\eps}}
\right)^{\eps} \\
&\leq \left( \sum_{\vec{k}} q_{\vec{k}} (c+1)^{\frac{1}{\eps}}
h_A(\tilde{S}_{\vec{k}}) \right)^{\eps} \nonumber\\
&=\left( (c+1)^{\frac{1}{\epsilon}} h_A(\src^{\otimes n}) \right)^{\eps} \nonumber\\
&\leq (c+1) \delta^{\eps} \nonumber
\end{align}
The inequality in the last line follows from the typical subspace
theorem. As $\delta$ can be chosen to be arbitrarily small, this
concludes our proof.

\end{proof}

\end{document}